\documentclass{article}

\usepackage{arxiv}

\usepackage[utf8]{inputenc} 
\usepackage[T1]{fontenc}    
\usepackage{hyperref}       
\usepackage{url}            
\usepackage{booktabs}       
\usepackage{amsfonts}       
\usepackage{nicefrac}       
\usepackage{microtype}      
\usepackage{lipsum}		
\usepackage{graphicx}
\usepackage{natbib}
\usepackage{doi}
\usepackage{amsmath}

\title{Defining Utility as a Measure of Preference Under Uncertainty in Phase I-II Oncology Dose Finding Trials}

\date{October, 2025}	

\author{
Andrew Hall \\
	Institute of Clinical Trials Research\\
	University of Leeds\\
	Leeds, United Kingdom \\
    \texttt{a.hall2@leeds.ac.uk} \\
	\AND
Duncan Wilson \\
	Institute of Clinical Trials Research\\
	University of Leeds\\
	Leeds, United Kingdom \\
\And
Stuart Barber \\
	School of Mathematics\\
	University of Leeds\\
	Leeds, United Kingdom \\
\And
    Sarah R Brown \\
	Institute of Clinical Trials Research\\
	University of Leeds\\
	Leeds, United Kingdom 
} 

\renewcommand{\shorttitle}{R2DT - Reference Dependent Decision Theoretic Dose Finding}

\hypersetup{
pdftitle={Defining Utility as a Measure of Preference Under Uncertainty in Phase I-II Oncology Dose Finding Trials},
pdfsubject={Phase I-II Oncology Dose Finding Trials},
pdfauthor={Andrew Hall, Duncan Wilson, Stuart Barber and Sarah R Brown},
pdfkeywords={Dose Finding, Phase I-II, Decision Theoretic, Reference Dependence, Utility},
}

\begin{document}
\maketitle

\begin{abstract}

The main objective of dose finding trials is to find an optimal dose amongst a candidate set for further research. The trial design in oncology proceeds in stages with a decision as to how to treat the next group of patients made at every stage until a final sample size is reached or the trial stopped early.

This work applies a Bayesian decision-theoretic approach to the problem, proposing a new utility function based on both efficacy and toxicity and grounded in von Neumann-Morgenstern (VNM) utility theory. Our proposed framework seeks to better capture real clinical judgements by allowing attitudes to risk to vary when the judgements are of gains or losses, which are defined with respect to an intermediate outcome known as a reference point. We call this method Reference Dependent Decision Theoretic dose finding (R2DT). 

A simulation study demonstrates that the framework can perform well and produce good operating characteristics. The simulation results demonstrate that R2DT is better at detecting the optimal dose in scenarios where candidate doses are around minimum acceptable efficacy and maximum acceptable toxicity thresholds.  

The proposed framework shows that a flexible utility function, which better captures clinician beliefs, can lead to trials with good operating characteristics, including a high probability of finding the optimal dose. Our work demonstrates proof-of-concept for this framework, which should be evaluated in a broader range of settings.
\end{abstract}


\section{Introduction} \label{sct:intro}
The dose selection paradigm in oncology has been shaped historically by the prognosis of diagnosis, the lack of effective treatments, and the properties of cytotoxic treatments coming through development \cite{FourieZirkelbach2022}. The effectiveness and toxicity associated with a cytotoxic agent increase steeply with increasing dose; the highest dose of a cytotoxic agent that patients can tolerate, based on a composite binary toxicity endpoint called dose-limiting toxicity, is considered optimal for progress to phase II testing \cite{Eisenhauer2015, Marshall2012}. In recent times, advances in oncology treatment have come more predominantly from targeted agents \cite{Mukherjee2011}. The optimal dose for a targeted treatment may no longer correspond with the maximum tolerated dose, and a measure of efficacy alongside toxicity may be required \cite{Strohbehn2023}. Project Optimus is an FDA initiative to reform the dose optimization and dose selection paradigm in oncology drug development in response to the increased proportion of targeted treatments coming into development \cite{FDA2023}. Part of the strategy is dose selection that not only limits the toxicity but contrasts this with the efficacy of a treatment. Dose-finding trials with objectives incorporating both an efficacy and toxicity endpoint can be referred to as phase I/II or phase I-II designs. 

The scientific objective of a dose-finding trial is to determine a dose for the treatment of patients in the future \cite{Healy1978}. The ethical objective is to ensure patients studied within the trial are not exposed to excessive toxicity or doses with minimal efficacy. Lower-level objectives are concerning efficiency and reliability; the trial should utilize the minimum number of patients and be capable of finding an optimal dose with a degree of statistical accuracy \cite{Eisenhauer2015}. To meet the ethical objective, a dose-finding design proceeds in stages so that patients are treated optimally according to accumulating evidence. This staged approach is prominent in oncology due to the serious nature of side effects associated with treatments, and due to the fact that patients who enter the trial are seeking a therapeutic advantage. There are a number of staged statistical trial designs that aim to meet trial objectives, and these can be classified into two categories: model-based and model-assisted designs \cite{Gooley1994, Hunsberger2005, Hardwick2003, Ananthakrishnan2018, Whitehead2004, Whitehead2005, Takeda2018, OQuigley2001, Braun2002, Thall2004, Yuan2016a}. One of the main components of a statistical design is how a dose is selected for the next group of patients at each stage and for further study; this is referred to as the decision process in this paper.

A Bayesian decision-theoretic approach is a statistical method to determine an optimal action from a set of possible alternatives when the outcome is uncertain \cite{French2000}. There are two main components: a Bayesian model representing the structure of a system and its associated uncertainty; and a consequence or utility function \cite{Smith2010}. Utility is a numerical measure of consequence that follows an axiomatic basis for rational decision making. A decision maker faced with a set of alternative choices acts optimally by selecting the alternative that maximizes the expected utility, provided the utility function follows the four axioms of Von Neumann–Morgenstern rationality: completeness, transitivity, continuity, and independence \cite{VonNeumann1947}. A fully decision-theoretic approach is scientifically sound, providing coherent decisions when each of the components can sufficiently be determined. The Bayesian decision analysis approach, usually referred to as a decision-theoretic design, has been applied to dose finding \cite{Whitehead1998, Whitehead2004, Whitehead2005, Thall2012, Yuan2016a, Loke2005, Wang2009a}.

The decision process in virtually all phase I trial designs utilising a model is similar to a Bayesian decision-theoretic approach with the feature of maximizing (or minimizing) some function measuring a payoff between efficacy and toxicity to choose a dose at each stage. This function has been referred to as an objective, loss, gain, value, or utility function without consistency. Additional ad hoc rules are typically imposed to meet the ethical objective for patients; these are included in the Bayesian decision-theoretic approaches also. The main ad-hoc rules, that are common place, are admissibility criteria to define an evidence level for an estimated minimum amount of efficacy and maximum amount of toxicity in order for a dose to be considered in the decision process. These admissibility rules have also been described as over-dose control and under-dose control \cite{Chen2015}. This initial restriction of the decision space through admissibility criteria departs from the principles of a fully Bayesian decision-theoretic approach, which assumes all doses are evaluated through the utility function alone. This paper aims to eliminate the reliance on such ad hoc constraints, instead embedding ethical considerations directly within a utility function that more closely reflects the clincial setting. This method is referred to as Reference Dependent Decision Theoretic dose finding (R2DT) from here on. 

The remainder of this section introduces a motivating example, highlights the importance of considering uncertainty in decision-making, which isn't considered in existing dose-finding designs, and defines the concept of \textit{attributes}, which are the measurable components used within the utility function.

\subsection{Motivating Example}
The motivation for the work in this paper came from designing a dose-finding study through the Leeds Institute of Clinical Trials Research. The study was in relapsed-refractory multiple myeloma, a cancer of the plasma cells, with the aim of investigating four doses of a treatment in combination with fixed-dose standard of care therapies. a phase I-II design was deemed appropriate by the clinical team, with the toxicity endpoint being a binary indicator of whether a dose-limiting toxicity is experienced in the first two four-week treatment cycles. The efficacy endpoint was also binary, recording whether or not the patient achieved a partial response within the same time period. 

\begin{figure}[ht!] 
\includegraphics[width=\textwidth]{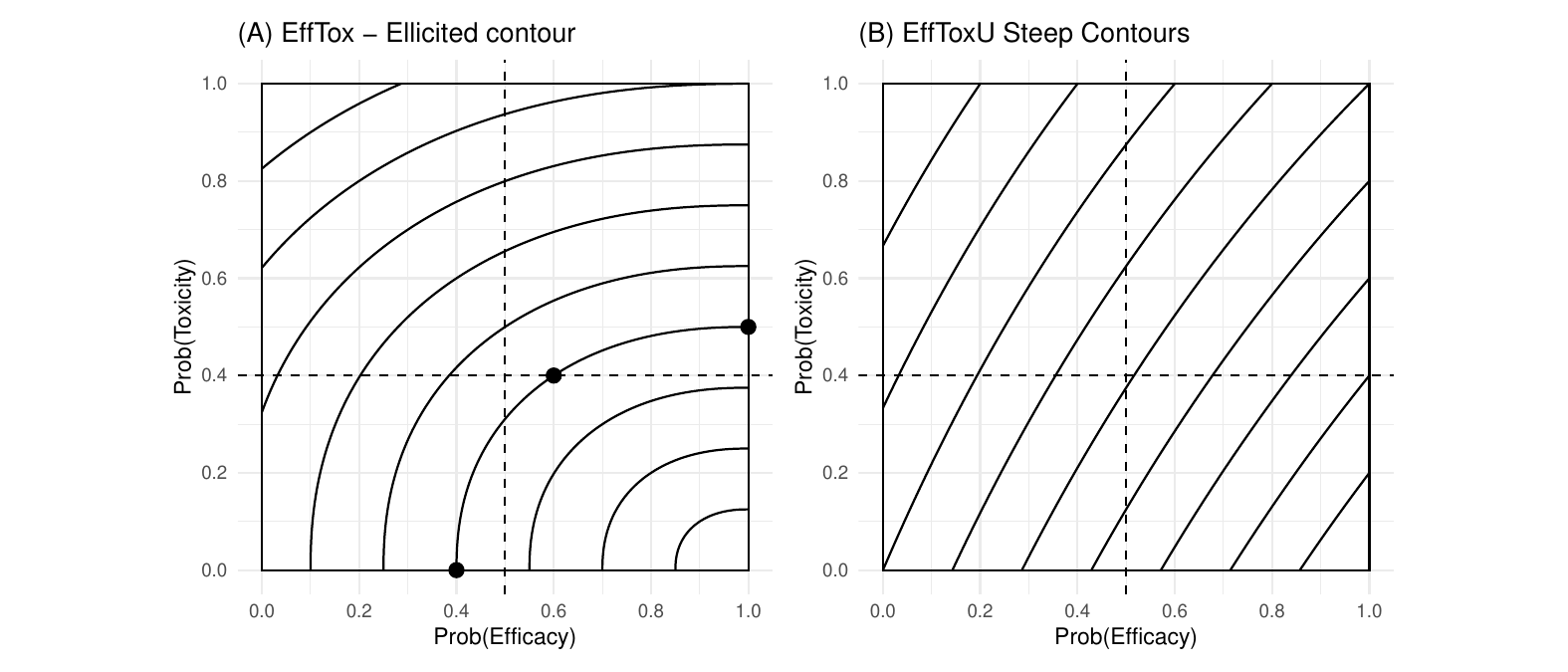}
   \caption[EffTox trade-off contours for motivating example]{(\textbf{A}): EffTox trade-off contours (solid lines) with elicited points $(0.4,0), (1,0.5)$ and $(0.6,0.4)$. Lines describe equal desirability, contours closer to bottom right corner (1,0) more desirable. Dashed lines represent admissibility criteria $\overline{\pi}_{addE}=0.5$ and $\overline{\pi}_{addT}=0.4$.
   (\textbf{B}): Utility EffTox contours (solid lines) are defined by elicited utility values: $0.5$ for a patient experiencing both efficacy and toxicity, and $0.3$ for a patient experiencing neither efficacy nor toxicity. Lines describe equal desirability, contours closer to bottom right corner (1,0) more desirable. Dashed lines represent admissibility criteria $\overline{\pi}_{addE}=0.5$ and $\overline{\pi}_{addT}=0.4$.}
    \label{fig:fig1}
    \end{figure}

The \textit{EffTox} \cite{Thall2004} design was considered. A subsequent (and recommended) iteration of the decision process converts the trade-off for efficacy and toxicity probability outcomes into one dimension by a set of vector norm contours, describing lines of equal desirability \cite{Cook2006, Brock2017}. The decision process after each cohort finds the mean estimate of parameters from the probability model to yield point estimates for the probability of efficacy and toxicity at each dose. A ranking is created, called a consequence function here, by the distance the contours are from the outcome with perfect efficacy without a chance of toxicity. In this example, the contour was informed by three elicited points in consultation with the clinical team, Figure \ref{fig:fig1}. Admissibility rules were also elicited to define four quadrants in the outcome domain, with the lower right quadrant deemed admissible.

The performance of the design was assessed by simulating a number of scenarios with different dose efficacy and dose toxicity relationships. These are called operating characteristics. The specified design performed poorly, with a tendency to get stuck at lower doses in scenarios where the highest and second-highest doses from the four doses were determined to be optimal. This is a known artifact of the design; authors of the \textit{EffTox} method stress the importance of contour specification, contours that are ``insufficiently steep'' will lead to ``pathological behaviour''. This is the tendency of a design to repeatedly recommend a low dose without exploring higher doses, that may be more optimal. From a visual inspection of the contours in Figure \ref{fig:fig1}A, the gradient of the right edge of the contours is near horizontal in contrast to the left edge. When the clinical team was asked to consider two points on any given contour in the admissible region, the team had a strong preference for doses towards the right-hand side of the contour. This suggests that the contours do not represent lines of equal preference. It would be possible to reformulate the questions so that the initially elicited contour would be contained in the lower right quadrant. In this instance, this would produce steeper contours, as in Figure \ref{fig:fig1}B, associated with improved operating characteristics.

Designs based upon a utility function to represent the merit of the four possible patient outcomes (efficacy alone, efficacy and toxicity, toxicity alone and neither efficacy or toxicity) have previously been proposed as an alternative to trade-off contours and have been described as numerical utilities \cite{Thall2012, Yuan2016a, Shi2021, Zhou2019}. This approach, labelled \textit{EffToxU}, is more akin to a Bayesian decision-theoretic approach than \textit{EffTox}, as the dose with the highest expected utility at each stage is chosen. It is reported that clinicians can easily comprehend the meaning of numeric utility and provide specifications that align with clinical judgments \cite{Zhang2024}. Figure \ref{fig:fig1}B gives such an example, with constants that are likely to work in many settings \cite{Yuan2016a}. Applying this design to the motivating example, when considering steep contours in the upper right quadrant, the clinical team had a strong preference for doses to the left of the contour where toxicity is lower and acceptable. The elicitation in both methods may be a reasonable reflection of the clinical preferences, but when extrapolating to create a function across the outcome domain, the functions are unable to be a reasonable approximation to the clinical situation. The consequence function and the numerical utilities specified by \textit{Efftox} and \textit{EfftoxU} respectively are a simplification of the situation described in the motivating example.

Many authors consider the consequence function as part of the statistical design with a set of components that need tuning through simulation, to give good operating characteristics, accompanied by less formal clinical consultation \cite{Thall2008}. Admissibility criteria are typically necessary components of trial design, compensating for a simplified consequence function and preventing unethical choices for patients. This is an important part of the design process, whereby the design will push for higher doses until there is sufficient evidence for the dose to be excluded \cite{Brock2017}. The level of evidence before initiating the admissibility rules can be lowered to better capture the objective of not exposing patients to excessive toxicity or doses with minimal efficacy, but this will lead to poorer design performance through pathological behaviour. When there is minimal evidence available, the admissibility rules are likely not efficient in excluding doses that appear to be quite toxic or not efficacious, due to the small number of patients. Statistically, the use of admissibility rules constitutes a two-stage approach to decision making, by restricting the doses under consideration according to the admissibility rules before maximising an objective function. The initial admissibility component of the design falls short of a fully decision-theoretic approach \cite{Wang2009a}. The next subsection will explore the importance uncertainty in decision making and defining a utility function. 

\subsection{The Importance of Uncertainty}
In a dose-finding trial, at each decision point, the probability of efficacy or toxicity isn't known precisely as little data exists. This is called uncertainty and is captured in the Bayesian paradigm by the posterior probability density function. Maximising the utility function with an inappropriate scale in the presence of uncertainty can lead to poor decision making. This was made famous by the so-called St. Petersburg paradox, first described in the 18th century and accompanied the early development of utility theory \cite{Bernoulli1954}. Preferences concerning a decision can be different when faced with an uncertain situation. 

Strength of preference refers to how strongly an individual prefers one option over another. It is a qualitative measure that indicates the intensity of preference between choices. For example, consider the chance of an efficacy event for a binary variable, for three doses A, B and C. The chance of success for a patient is 50\%, 75\% and 100\% for each dose respectively. It is clear in all clinical settings that the three doses can be ranked. It is not possible to deduce however whether the 25\% increase in efficacy between doses A and B is preferred more than between doses B and C, from the percentages; with the interpretation assumed to change for different clinical settings. In the example, the raw chance of experiencing an efficacy event represents a numerical ranking; it is on an ordinal scale rather than interval. It is important to have an interval scale so that differences in preference strength can be meaningfully interpreted and compared, especially when evaluating trade-offs between efficacy and toxicity. Trying to obtain a ``strength of preference'' measure so that a utility function can be defined on a interval scale is not an idea that is easily articulated or elicited \cite{French2000}. The Von Neumann–Morgenstern (VNM) utility theorem overcomes the difficulties by considering the problem in terms of preferences for lotteries described by probability density functions from the outset, so that a utility function can be defined on an interval scale \cite{VonNeumann1947}. 

Inspecting the \textit{EffToxU} design implies indifference to decisions under uncertainty for efficacy and toxicity (assuming an independence probability model, Appendix \ref{app:EffToxU}). A lottery is able to describe how preferences in an uncertain situation relate to a certain situation. For example, given any fixed toxicity rate, the \textit{EffToxU} utility function would suggest indifference to receiving a treatment with certain 75\% efficacy or facing a 50-50 lottery between a treatment with 100\% efficacy and 50\% efficacy. Importantly, this is true for any specification of the numerical utility values. 

A utility function that is able to capture different preferences for uncertainty would allow us to account for different levels of acceptable patient risk, applicable to different clinical settings, while simultaneously defining our utility function on an interval scale. von Neumann-Morgenstern utility theory is foundational to Bayesian decision theory, providing the justification of maximising expected utility. The interpretation of utility in VNM is made with respect to uncertainty; something that is inherent in dose finding as a result of the small number of patients. To our knowledge, no previous design in the Phase I-II dose-finding literature has been defined with respect to the VNM utility theory by considering preferences under uncertainty.  

\subsection{Attributes}

In decision making, objectives are characterised by \textit{attributes}; these are measures that are used in the utility function. The decision for which dose to choose at the end of the trial has implications for a population of patients. Interim decisions part way through the trial also typically have implications for more than one patient (although often only three, based upon the typical cohort size), which belong to this same patient population. As such, attributes for R2DT are the population parameters for the probabilities that a patient will experience an efficacy and a toxicity event. Admissibility criteria in the literature define a threshold to split each population level attribute into regions of acceptability or unacceptability. R2DT considers attributes for toxicity and efficacy against a similar reference point, that changes depending on the clinical setting; the merit of an incremental increase in either attribute is considered differently depending on whether it is considered a ``gain'' upon the reference or a ``loss''. The distance from the reference point is also a factor. Framing each attribute with respect to the reference point is called reference dependence. Creating a utility function for each attribute through elicitation of uncertain outcomes and considering reference dependence allows us to incorporate the ethical objectives directly into consideration of the optimal dose rather than with separate admissibility criteria.

\subsection{Structure of the Paper}
This paper proposes a framework that seeks to more closely capture a given clinical setting through the Bayesian decision-theoretic approach. It captures uncertainty by following the axiomatic basis of VNM utility theory that is elicited through clinical input. 

The paper is structured as follows: The overarching trial design and the Bayesian decision theoretic approach is restated for the setting before a closer inspection of defining utility functions based upon reference dependence and attitudes to uncertainty for efficacy and toxicity attributes separately. Multivariate utility theory gives a broad form for the utility function with constants to be set according to clinical input. An elicitation protocol is given for the method to define all parameters in a utility function. The merits of the method are then evaluated using simulation, linking to the initial motivating example.

\section{Methods}

\subsection{Bayesian Decision Theory in Dose Finding}
The following section introduces the general framework for the Bayesian decision theoretic approach used by R2DT. 

Let $D = \{d_1<d_2<\cdots<d_k\}$, where $d_j \in \mathbb{R}_{>0}$, be a set of $k$ pre-defined doses to be studied and $Y=(Y_E,Y_T)$ where
\begin{equation}
	\begin{split}
		Y_E=\begin{cases}
		1 & \text{if efficacy}  \\
		0 & \text{otherwise} \\
		\end{cases}
	\end{split}   
\qquad \text{and} \qquad
	 \begin{split}
		Y_T=\begin{cases}
		1 & \text{if toxicity}  \\
		0 & \text{otherwise} \\
		\end{cases}
	\end{split}
\end{equation}
are Bernoulli random variables representing an efficacy and toxicity event respectively. Each event definition will depend on the particular clinical setting. For example, efficacy may be measured by response or progression-free survival at a particular time point. We assume that both efficacy and toxicity endpoints are measured over a similar time period. Features that are unknown about the external world, namely the probability of efficacy and toxicity at each dose are modelled by a vector of parameters, $\theta \in \Theta$ to denote unknown states of nature. The observation $Y$ is drawn from a distribution $ p_Y(y|\theta)$. Prior knowledge of $\theta \in \Theta$ is incorporated via a prior $p_\theta(\cdot)$.
This is updated through Bayes theorem in light of the observation(s), to give the posterior
\begin{equation}
    p(\theta|y) \propto   p(y|\theta) \times p(\theta).
\end{equation}
The probability model $p(y|\theta)$ for the R2DT method follows independent logistic regression models for efficacy and toxicity with normal priors for regression coefficients, similar to previous work in this setting \cite{Thall2004}:
The covariate for a dose $d \in D$ is transformed by centering around the geometric mean; that is,
\begin{equation}
f(d) = \log(d)-\frac{1}{k}\sum\limits_{r=1}^{k} \log(d_r). 
\end{equation} 
An inverse-logit link function is used to relate probabilities of efficacy at dose $d$, denoted by $\pi_{E}$, and of toxicity, denoted by $\pi_{T}$:
\begin{equation}
\pi_{E} = \text{logit}^{-1}\{\mu_E + \beta_{E1}f(d) + \beta_{E2}[f(d)]^2\} 
\end{equation}
\begin{equation}
\pi_{T} = \text{logit}^{-1}\{ \mu_T + \beta_Tf(d)\}. 
\end{equation}
The additional squared term in the efficacy model, with coefficient $\beta_{E2}$, allows for the possibility that efficacy may not be monotonic in dose. Model parameters for the design are defined by $\boldsymbol\theta= (\mu_E, \beta_{E1}, \beta_{E2}, \mu_T, \beta_T)$ and data for a patient $i$ by $\mathcal{D}_i=(Y_i,z_i)$. Prior distributions for individual model parameters follow an independent normal distribution with corresponding hyper parameters for the mean and variance.

A utility function $u(\pi_{E},\pi_{T})$ specifies the utility of treating when the probabilities of efficacy and toxicity are at $\pi_E$ and $\pi_T$. These probabilities are determined by the states of nature is $\theta$ and dose $d \in D$. At each stage, the potential actions are selecting a dose $d$ to treat the next cohort. The Bayes action (or decision) $d^* \in D$ is the action that maximises the posterior expected utility: 
\begin{equation}
    d^*(y) = \underset{d}{\arg\max}( E [u(\pi_{E},\pi_{T})|y,d] ).
\end{equation}
The individual or group responsible for agreeing on the utility function, in consultation with the statistician, is referred to as the Decision Maker (DM) in this paper. The trial recruits in cohorts of size $c$ with the posterior formed from data after each cohort. The Bayes decision determines the dose for the next cohort. No skipping of untried doses in escalation is stipulated as an additional safety rule outside of the probability model to account for model misspecification in earlier cohorts \cite{Thall2004}. Specifically, if the Bayes decision is more than one dose above the highest dose previously assessed, then the dose for the next cohort will be the dose one level above the highest dose previously assessed.  The trial continues until a maximum sample size is reached, with the Bayes decision following the final cohort determining the dose suggested for subsequent research.    
 
\subsection{R2DT Utility Specification}

If we accept the axioms of utility theory, there exists some true, non-parametric, utility function which could (at least in principle), be determined through some elicitation process. For utility functions with two attributes, this elicitation can be challenging \cite{Keeney1993}. In R2DT we overcome this by first inspecting the preference structure  to yield a broad family of parametric utility functions, which can then be fully specified according to clinical input. 

R2DT has a utility function of the form $u(\pi_E,\pi_T)=f(u_E,u_T)$ with $f(\cdot)$ a simple function, $u_E$ a marginal utility function of $\pi_E$ (given any value of $\pi_T$) and $u_T$ a marginal utility function of $\pi_T$ (given any value of $\pi_E$). In doing so, more easily assessed marginal utility functions can be formed before considering the more complicated bivariate form. 

The rest of this section is structured as follows: The marginal utilities are first defined with attitudes to uncertainty and reference dependence. The two functions are combined in Section \ref{sct:util_ind} accounting for how the two utilities interact. An extension to R2DT looks at the role of utility in stopping the trial in light of all doses being overly toxic and/or efficacious, expanded upon in Section \ref{sct:stopping_rules}. Finally an elicitation protocol is described to obtain all parameters of R2DT.  

\subsubsection{Marginal Utility Functions}
Utility functions capture preferences under uncertainty, which can be described by simple lotteries. Consider a scenario where the DM faces an outcome $\pi_E=x_1$ with probability $\alpha$, or $\pi_E=x_3$ with probability $1 - \alpha$. We refer to this as a lottery between $x_1$ and $x_2$ with a mixing component of $\alpha$, and denote it by  $\langle x_1, \alpha, x_3\rangle$.  The relation is abbreviated to $\langle x_1,x_3\rangle$ when denoting an equal lottery with $\alpha=0.5$. We use the notation $a \sim b$ to denote the case where the DM is indifferent between outcomes $a$ and $b$. For example, the statement 
  \begin{equation} \label{eqn:lottery_notation}
    \langle x_1,\alpha,x_3\rangle ~ \sim ~ x_2
\end{equation}
tells us the DM is indifferent between either receiving the lottery between $x_1$ and $x_3$ or the certain outcome $\pi_E=x_2$.
For a utility function which follows VNM theory to reflect indifference, we require that
\begin{equation} \label{eqn:utility} 
\alpha u(x_1) + (1-\alpha) u(x_3) = u(x_2). 
\end{equation}

It is a self-evident principle that the utility function for efficacy is monotonically increasing and the toxicity utility function is monotonically decreasing, since more efficacy is invariably preferred to less and less toxicity is always preferred to more. If $x_1=\pi_E^*$ and $x_3=\pi_E^{**}$ are two levels of the efficacy attribute then 
\begin{equation}
    [\pi_E^*> \pi_E^{**}] \Leftrightarrow[u(\pi_E^*)>u(\pi_E^{**})]. 
\end{equation}
Similarly if $y_1=\pi_T^*$ and $y_2=\pi_T^{**}$ are two levels of the toxicity attribute then 
\begin{equation} \label{eq:monotox}
    [\pi_T^*>\pi_T^{**}] \Leftrightarrow[u(\pi_T^*)<u(\pi_T^{**})]. 
\end{equation}

\begin{figure}[ht!] 
\includegraphics[width=\textwidth]{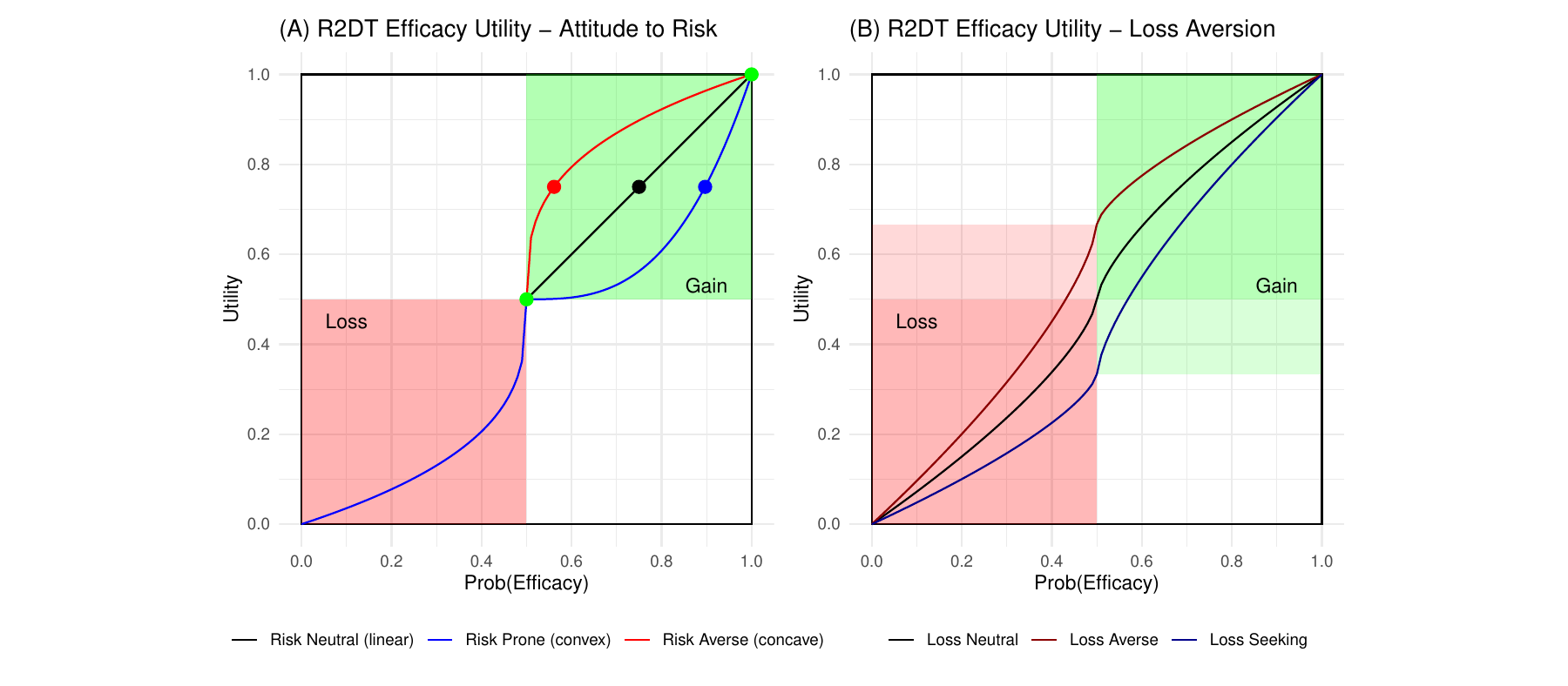}
   \caption[Attitudes to risk and loss aversion for efficacy utility function]{Attitudes to risk and loss aversion for efficacy utility function
    (A): Constantly increasing utility for efficacy attribute (probability of efficacy event) with reference probability of 0.5 defining whether a ``Loss'' or ``Gain''. In Gain domain, Green points represent the simple lottery $\langle x_1=0.5, x_3=1\rangle  \sim \underline{x_2}$, with $\underline{x_2}$ the certainty equivalent. When the DM is risk neutral, the certain equivalent is equal to the expectation of the lottery (black point, $\underline{x_2}=0.75$). Risk aversion is when the certainty equivalent is less than the expected consequence of the lottery (red point, $\underline{x_2}\le 0.75$). Risk prone is when the certainty equivalent is more than expected consequence of the lottery (bluepoint $\underline{x_2}\ge 0.75$). In Loss domain, R2DT proposes a risk prone function.     
    (B) Loss aversion is depicted with an example sigmoid efficacy utility function proposed for R2DT (convex for ``Losses'' and concave for ``Gains'). Loss aversion is specified to reflect ethical objective of avoiding exposure to non-efficacious doses and has the effect of stretching the loss region}
    \label{fig:fig2}
    \end{figure}

A utility function reflects a decision maker's preferences when faced with uncertain options, and this will be influenced by their attitude to risk. Risk aversion is described as a preference for the expected consequence of a lottery over the lottery itself, while risk prone is a preference for the lottery and risk neutrality is indifference between the two. For example, consider a lottery between three doses. doses 1 and 3 have an efficacy of $\pi_{E,1} = 0.5$ and $\pi_{E,3} = 1$. When choosing between a simple lottery between these two doses and a certain outcome of $\pi_{E,2}$, risk neutrality would correspond to $\pi_{E,2} = 0.75$. A risk prone DM would have $\pi_{E,2} > 0.75$, while a risk averse DM would have $\pi_{E,2} < 0.75$. For a parametric increasing utility function where a DM has a consistent attitude to risk, a concave function describes risk aversion, a convex function describes risk prone, and a linear function describes risk neutrality \cite{Keeney1993}. Three utility functions illustrating these attitudes are given in the top right of Figure \ref{fig:fig2} (A).  Lotteries are used to elicit parameter choices as part of the parametric utility function defined in R2DT.  

A key feature of R2DT is reference dependence, where reference points are used to frame attitudes to risk for both attributes. Reference points are denoted by $\overline{\pi}_E$ and $\overline{\pi}_T$ for efficacy and toxicity respectively. The efficacy reference, $\overline{\pi}_E$, is suggested to correspond with the current efficacy estimates for standard of care rather than an aspirational level associated with the continued development of the drug. The toxicity reference, $\overline{\pi}_T$ is suggested to be thought as a target toxicity level, typically associated with toxicity-only dose finding designs \cite{Wheeler2019}. The attributes for efficacy and toxicity at each dose are transformed as follows $\pi_{E} - \overline{\pi}_E$ and $\overline{\pi}_T-\pi_{T}$ respectively. Note the transformation of toxicity attribute to satisfy Equation \ref{eq:monotox} ($[1-\pi_{T}] - [1 - \overline{\pi}_T]=\overline{\pi}_T-\pi_{T}$). A negative transformed attribute is labelled a ``loss'' and a positive value a ``gain'' upon the reference. For example, an improvement from the reference for efficacy is perceived to be beneficial for the patient i.e a ``gain'', with $\pi_{E} - \overline{\pi}_E > 0 $.   

Considering either attribute as reference dependent, attitudes to risk are assumed to be dependent upon whether considering the level of the attribute as a gain or a loss. Both marginal utility functions for R2DT are defined using a piece-wise function that splits the attribute into gains and losses. The power function is specified for each segment of the utility functions, as it is a parametric utility function where the attitude to risk depends upon the distance from the reference point \cite{Pratt1964}. The power utility is also a commonly used utility function when an attribute is measured relative to a reference \cite{Keeney1993}.

The following segmented power utility function is proposed for efficacy:  
\begin{equation} \label{eq:pow_ind_util_eff}
u_E(\pi_E) = \begin{cases}
                  g\bigl((\pi_E  - \overline{\pi}_E)^{\alpha_{GE}}\bigr)  & \pi_E \ge \overline{\pi}_E \\
                  g\bigl(-\lambda_E |\pi_E  - \overline{\pi}_E | ^ {\alpha_{LE}}\bigr) & \pi_E < \overline{\pi}_E,  \\
                \end{cases}
\end{equation}
 with $\lambda_E \ge 0$, $\alpha_{GE} \ge 0$, $\alpha_{LE} \ge 0$ and  $g(u) = [u - u(0)]/[u(1)-u(0)]$. where $u(1)$ and $u(0)$ are the maximum and minimum of the utility function at the points $\pi_E=1$ and $\pi_E=0$ respectively. The normalising function, $g$, scales the utility function to be in the range $[0, 1]$. The scaling is necessary to ensure the utility is on the same scale as the toxicity utility function when combining in Section \ref{sct:util_ind}. 
 
The parameters $\alpha_{LE}$  and $\alpha_{GE}$ specify the attitude to risk for losses and gains respectively; $\alpha_\cdot=1$ would indicate risk neutrality. We would suggest that $\alpha_{GE}<1$; this gives a concave (risk-averse) utility function for the gain segment. It is also proposed that $\alpha_{LE}<1$, representing a convex (risk-prone) utility function for losses. This may seem counter-intuitive at first, but if we were to reframe the outcomes relative to the reference point, the DM is effectively acting in a risk-averse manner as outcomes worsen. With the extreme values of $\alpha_{GE} =  0$ and $\alpha_{LE} = 0$, the utility function becomes a step function.  

The loss aversion index, $\lambda_E$, considers the merit of ``gains'' with respect to ``losses'' (Figure \ref{fig:fig2}(B)). Loss neutral, $\lambda_E=1$, considers gains and losses as equally important.  Increasing the loss aversion index so that $\lambda_E>1$ represent an increasing preference of avoiding losses more so than pursuing gains. In dose finding, loss aversion corresponds with the ethical objective of avoiding exposing patients to in-efficacious doses. The shape of the efficacy utility is sigmoidal as depicted in Figure \ref{fig:fig2}(B). 

The following utility function is proposed for toxicity.    
\begin{equation} \label{eq:ind_util_tox}
u_T(\pi_T)=         \begin{cases}
                 h\bigl((\overline{\pi}_T - \pi_T)^{\alpha_{GT}} \bigr) & \pi_T \le \overline{\pi}_T\\
                 h\bigl(-\lambda_T|\overline{\pi}_T - \pi_T|^{\alpha_{LT}} \bigr) & \pi_T > \overline{\pi}_T  \\
                 
                \end{cases}
\end{equation}
 with $\lambda_T \ge 0$, $\alpha_{GT} \ge 0$, $\alpha_{LT} \ge 0$ and  $h(u) = [u - u(1)]/[u(0)-u(1)]$. The normalising function, $h$,  places the utility function in the range $[0, 1]$. It is proposed that $\alpha_{GT}<1$, $\alpha_{LT}<1$ and $\lambda_T>1$ with similar interpretation and attitudes to risk to the efficacy utility function. Due to the initial transformation of the attribute the toxicity utility mirrors the efficacy utility i.e. an inverted sigmoidal shape. 

\subsubsection{Utility Independence} \label{sct:util_ind}
Utility functions for efficacy and toxicity are assumed to be mutually utility independent, Appendix \ref{app:joint_utility}, with both marginal utility functions defined in the range $[0, 1]$.  This gives the following joint utility \cite{Keeney1971}:
\begin{equation} \label{eq:utili_indep}
u(\pi_E,\pi_T) = k_E~u_E(\pi_E) + k_T~u_T(\pi_T) + k_{ET}~u_E(\pi_E)~u_T(\pi_T)
\end{equation}
 where
\begin{enumerate}
\item$0\le k_E \le 1$ and $0\le k_T \le 1$, 
\item $u_E$ is a marginal utility function for $E$,
\item $u_T$ is a marginal utility function for $T$,  
\item $k_{ET}= 1 - k_E - k_T$.
\end{enumerate} 
The marginal utility functions $u_E$ and $u_T$ have been established in the preceding subsections, leaving the two parameters $k_E$ and $k_T$ to be determined. $k_E$ defines the utility when $u_T(\pi_T)=0$ and $u_E(\pi_E)=1$ (the point $\pi_E=\pi_T=1$) and $k_T$ defines the utility when $u_T(\pi_T)=1$ and $u_E(\pi_E)=0$ (the point $\pi_E=\pi_T=0$). The constant $k_{ET}$ represents an interaction between the two attributes. A smaller sum of $k_E$ and $k_T$ would constitute a greater interaction, while $k_{ET}=0$ would imply no interaction. A positive interaction, $k_E+k_T<1$, is proposed for R2DT (Appendix \ref{app:joint_utility}). 

\subsection{Stopping the Trial} \label{sct:stopping_rules}
Within the Bayesian decision theoretic framework, we used $d^*$ to denote the action which maximises expected utility at the decision point. In a dose finding trial the potential actions are to treat the next cohort of patients at a dose $d\in D$ in addition to the action to stop the trial early due to either a lack of efficacy across all doses or excessive toxicity. The action to stop the trial could be specified using an additional utility function that incorporates the attributes $\pi_E$ and $\pi_T$, or further attributes more akin to phase II considerations \cite{Stallard1999}. However, specifying such a utility would be challenging. We propose limiting the use of maximum expected utility to finding the optimal dose, and implement a separate criterion to decide if the optimal action is actually to stop the trial.

The following criterion based upon the R2DT utility function is proposed: 
\begin{equation} \label{eq:R2DTstp}
  \text{Pr}\{u(\pi_E,\pi_T) < u(\pi_E=\overline{\pi}_{UaddE},\pi_T=\overline{\pi}_{UaddT})\} > 1 - p_u.
\end{equation}
where $\overline{\pi}_{UaddE}$ and $\overline{\pi}_{UaddT}$ are constants to define a single point on a contour of acceptability from the R2DT utility function and $1-p_u$ a predefined threshold. If all doses $d\in D$ surpass the threshold then the optimal decision is stop the trial without a dose selected.  

\subsection{Elicitation}
This section outlines a series of steps for eliciting suitable values for the parameters in the R2DT utility function and associated stopping rules. These values are obtained by posing a set of precise preference-based questions involving simple lotteries, as defined in Equation \ref{eqn:lottery_notation}. An \underline{underline} in notation is used to denote the object being elicited. The basic approach is to fix all but one of the constants and determine the value that satisfies the specified relation.

To specify paramters in the R2DT efficacy utility function, the first task is to obtain $\overline{\pi}_E$, the reference point for efficacy. This parameter doesn't need a lottery to be established and a suitable question would be ``At what efficacy level is the current standard of care?''. To find $\alpha_{GE}$, restrict lotteries to values above the previously elicited reference point, i.e. $x_1 \ge \overline{\pi}_E$.
For example if $\overline{\pi}_E=40\%$,
\begin{equation}
    \langle \overline{\pi}_E ,\overline{\pi}_E+20\%\rangle \sim \underline{x_{2}}.
\end{equation}
could be used and an example question may be ``what level of efficacy would you be indifferent to receiving with certainty, compared to a 50-50 lottery between a treatment with 40\% efficacy and 60\% efficacy?''. It is expected that $x_2$ would be less than the expectation, $50\%$ to reflect the risk averse attitude. The segment of the utility function when efficacy is higher than $\overline{\pi}_E$ (a gain) is given in Equation \ref{eq:pow_ind_util_eff} which is substituted into Equation \ref{eqn:utility} and solved to find $\alpha_{GE}$. To elicit $\alpha_{LE}$ the same procedure is followed although the lottery is exclusively in the loss domain. In order to elicit $\lambda_E$, the value of $x_2$ is sought in a lottery with  $x_1<\overline{\pi}_E$ and $x_3>\overline{\pi}_E$. 

For the toxicity utility function the reference point $\overline{\pi}_T$ is specified as a target toxicity, corresponding with target toxicity levels that are specified in phase I toxicity-only designs \cite{Wheeler2019}. The three required parameters follow the same procedure as the efficacy utility function with separate lotteries in the gain, loss and then a mixed lottery. 

There are two parameters, $k_E$ and $k_T$ from Equation \ref{eq:utili_indep} to determine the joint distribution. If any two points $(\pi_E=x_1, \pi_T=y_1)$ and $(\pi_E=x_2, \pi_T=y_2)$ in the outcome domain are considered equivalent then the utility must also be equal, i.e. $u(x_1, y_1)=u(x_2,y_2)$. Obtaining two equivalences would yield two equations which could be solved simultaneously to obtain the two parameters. 

The R2DT stopping rule involves specifying a threshold contour beyond which any treatment with a lower utility would be considered unacceptable. Given that the full elicitation of the parametric utility has already happened all that is needed for this elicitation is a single point on the contour. The single point $(x,y)$ corresponds with constants $\overline{\pi}_{UaddE}=x$ and $\overline{\pi}_{UaddT}=y$ given in equation \ref{eq:R2DTstp}. One way of asking this is to consider an efficacy level that is seen as both feasible and constitutes a significant step in improving outcomes for patients. The question is then: what is the maximum amount of toxicity that would be considered acceptable for this level of efficacy?    

The elicitation methods described in this section give the minimum number of simple lotteries or points of indifference that need to be elicited in order to obtain the parameters of the utility function and the stopping rule. The utility function by definition of continuous attributes implies an infinite number of other possible simple lotteries within the joint attribute space; some of these should be checked to ensure consistency and revision of earlier elicitations may be required.

\section{Simulation}
The merits of the R2DT design are explored utilising simulation with comparison against \textit{EffToxU}, a utility design based upon eliciting the merit of a patient experiencing both toxicity and efficacy events or neither (Appendix \ref{app:EffToxU}) \cite{Thall2012, Yuan2016a}. In the comparator, the interaction between admissibility rules and the decision function plays an important role. R2DT proposes a single trial stopping rule, whereas admissibility criteria are central to the decision-making process in EffToxU. To account for this, the evaluation of R2DT follows a staged approach. It is first applied using conventional admissibility criteria (Appendix \ref{app:stopping_rules}) and compared against the comparator. This allows for a clearer understanding of the contributions of both the stopping rule and the utility function. The utility-based trial stopping rule is then evaluated as described in Section \ref{sct:stopping_rules}.

The designs are applied to a fictitious example in primary double-refractory multiple myeloma reflecting the motivating example. The toxicity endpoint in this setting is a binary indicator of whether a dose limiting toxicity is experienced in the first two, four-week cycles. Efficacy will be a binary variable as to whether the patient achieved a ``partial response'' within the same time period. The trial will investigate 4 doses of an investigational medicinal product with units mg/kg, $D=(20,30,40,50)$. 

Fixed probability vectors $\tilde{\pi}_E(D)$ and $\tilde{\pi}_T(D)$ are specified for 10 clinically plausible scenarios (Table \ref{tab:Table_2}, see also Appendix \ref{fig:scen_1_R2DT} and \ref{fig:scen_2_R2DT}). Simulated data is generated for each scenario for all patients at each dose for 2000 repeated trials. Outcomes for dummy patients are drawn according to $Y_E\ \sim B(\tilde{\pi}_E(d_j))$  and $Y_T\ \sim B(\tilde{\pi}_T(d_j))$ where $B$ is a Bernoulli distribution. Different trial designs are applied to the simulated data with the performance of designs assessed by operating characteristics, defined by the percentage of selection across the 2000 replicates and the average number of patients treated at each dose. The different trial designs are described in subsequent text and listed in Table \ref{tab:Table_1}. Utility contour plots for each design are plotted in Figure \ref{fig:quad_util} and Appendix \ref{fig:scenerio_utility}. 

The trial will start at the 20mg/kg dose. Successive cohorts of size $c=3$ will be recruited to the trial until the pre-defined maximum sample size of 45 is achieved or the trial stopped early. The impact of the R2DT design on sample size is explored as part of the simulation study. The patient group is expected to have a 50\% response rate if treated outside of the trial with the standard of care established agent. The established agent is generally well tolerated, with a target toxicity rate of 35\% for the agent under evaluation. 

The same probability model and priors have been specified for each of the different designs (Appendix \ref{tab:model_params}). Efficacy and toxicity are modelled independently. Priors have been specified according to a mean vector at each dose and equivalent sample size (ESS) \cite{Thall2014a}. The mean vector was chosen as the mean of the first six scenarios; a range of ESS values were explored for the \textit{EffTox} design utilising \textit{EffTox} software \cite{Thall2021}. The chosen ESS gave suitable operating characteristics across all scenarios. 

\subsection{R2DT simulation}
Table \ref{tab:Table_1} summarises each design in the simulation study.
\begin{table}[ht]
\caption{\label{tab:Table_1}Short description of each of the different methods in simulation study}
\centering
\begin{tabular}{p{0.25\linewidth} | p{0.7\linewidth}}
\toprule
Label & Description\\
\midrule
\multicolumn{2}{l}{\textbf{Method Comparison}} \vspace{0.2cm}\\
R2DT (1) & Sigmoidal and inverted sigmoidal shaped efficacy and toxicity marginal utility functions respectively. Joint utility $k_E=0.25$ and $k_T=0.15$. Admissibility rules applied as separate step functions at each dose. All parameters specified in Table \ref{tab:Table_2}.  \vspace{0.3cm}
 \\
EffToxU (2) & Marginal utilities are linear. Joint utility and admissibility rules applied as R2DT (1) \vspace{0.3cm}\\
\multicolumn{2}{l}{\textbf{R2DT Stopping Rule}} \vspace{0.2cm}\\
R2DT (3i) & R2DT (1) but single admissibility rule to limit doses under evaluation at each stage based upon contour, u(0.5,0.35)= 0.58\vspace{0.3cm}\\
R2DT (3ii) & as (i) but contour includes u(0.7,0.4)= 0.62\vspace{0.3cm}\\
R2DT (3iii) & as (i) but contour includes u(0.9,0.4)= 0.69\vspace{0.3cm}\\
R2DT (4i) & R2DT (1) but single trial stopping rule based upon u(0.5,0.35)= 0.58. All doses considered at each stage but trial stops if all doses considiered unacceptable \vspace{0.3cm}\\
R2DT (4ii) & as (i) but contour includes U(0.7,0.4)= 0.62\vspace{0.3cm}\\
R2DT (4iii) & as (i) but contour includes U(0.9,0.4)= 0.69\vspace{0.3cm}\\
\multicolumn{2}{l}{\textbf{Comparator Sensitivity}} \vspace{0.2cm}\\
EffToxU (5) & EffToxU (2) single admissibility rule based upon u(0.5,0.35)= 0.42\vspace{0.3cm}\\
EffTox (6) &  EffTox method applied defined from equal contour passing u(0.5,0.35)= 0.42. Admissibility rules applied as separate step functions at each dose\vspace{0.3cm}\\
EffToxU (7) &  EffToxU (2) but with $k_E=0.5$ and $k_T=0.3$ \\
\bottomrule
\end{tabular}
\end{table}

\subsubsection{Method Comparison}
\begin{figure}[ht!]
\includegraphics[width=\textwidth]{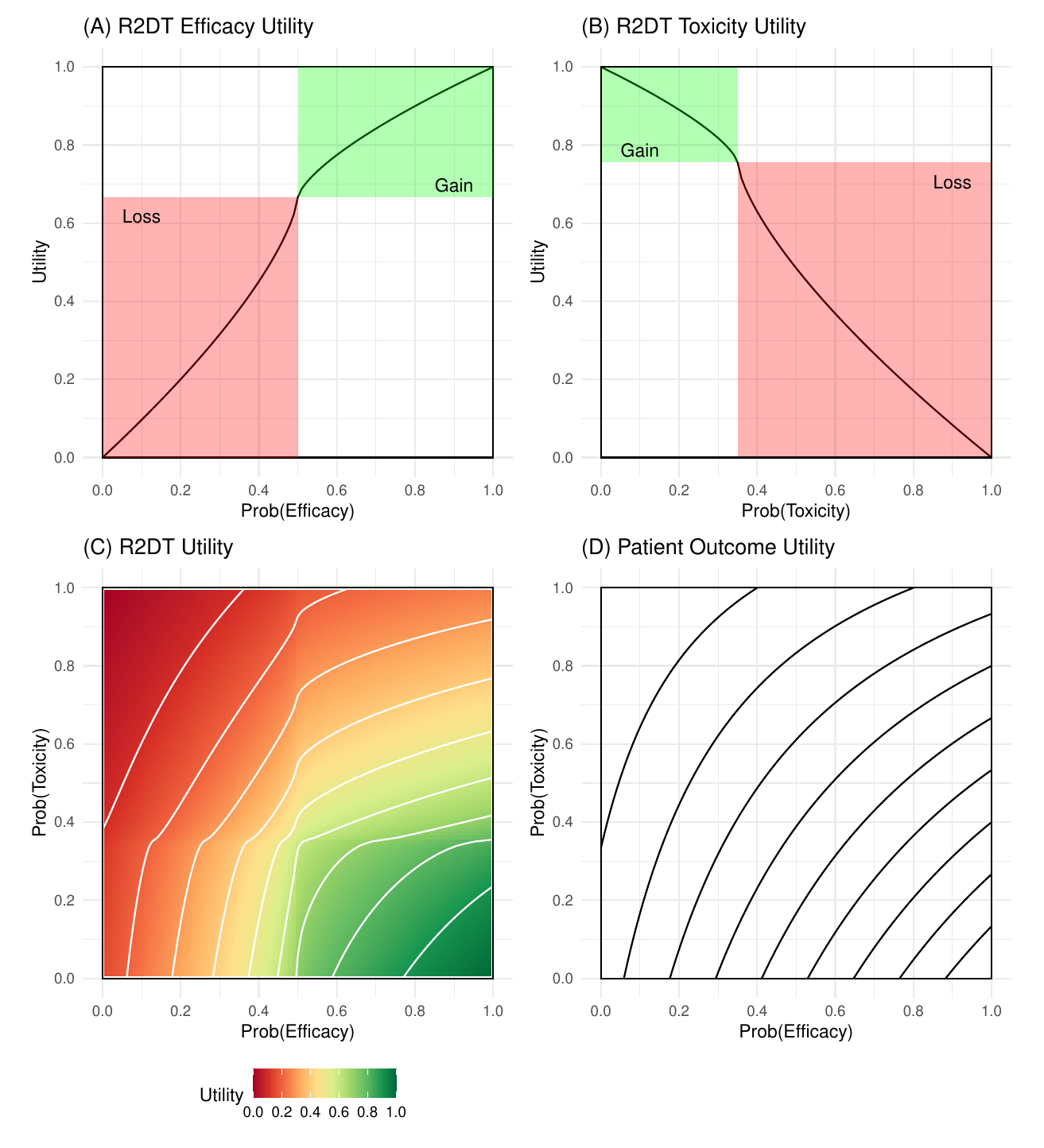}
  \caption{R2DT Utility function:  A,B,C depict R2DT (1) method in simulation study. D depicts joint utility function of \textit{EffToxU (2)}. Contours in the joint utility represent equal utility at 0.1,0.2,...,0.9 with the the point at guaranteed efficacy and no toxicity having utility of 1.}
     \label{fig:quad_util}
    \end{figure}

The initial comparison is between the R2DT method, labelled \textit{R2DT (1)}, and the \textit{EffTox} patient outcome utility design, labelled \textit{EffToxU (2)} with both designs having conventional admissibility criteria (Appendix \ref{app:stopping_rules}). Contour plots for \textit{R2DT (1)} and \textit{EffToxU (2)} are plotted in Figure \ref{fig:quad_util} and described in proceeding paragraphs. 

\textit{R2DT (1)} is specified using the marginal efficacy function, marginal toxicity function and joint utility function for R2DT. The marginal toxicity utility is determined by Equation \ref{eq:ind_util_tox} with parameters as specified in Table \ref{tab:utilitit_ params} and plotted in Figure \ref{fig:quad_util}A. The marginal toxicity utility is determined by Equation \ref{eq:ind_util_tox} with parameters as specified in Table \ref{tab:utilitit_ params} and plotted in Figure \ref{fig:quad_util}B. The joint utility combines the two marginal utility functions following Equation \ref{eq:utili_indep}, with constants specified in Table \ref{tab:utilitit_ params}. The joint utility function is plotted in Figure \ref{fig:quad_util}C. Admissibility rules for efficacy and toxicity are applied as per Equations \ref{eq:admis_e} and \ref{eq:admis_t} (Appendix \ref{app:stopping_rules}), with $\overline{\pi}_{addE}=0.5$, $p_E=0.075$, $\overline{\pi}_{addT}=0.4$ and $p_T=0.075$. 

\begin{table}[ht!]
\caption[Short description of each of the different constants and their interpretation specified in simulated R2DT utility function]{\label{tab:utilitit_ params}Short description of each of the different constants and interpretation specified in \textit{R2DT (1)}}
\centering
\begin{tabular}{p{0.25\linewidth} | p{0.7\linewidth}}
\toprule
Constant & Description\\
\midrule  
\multicolumn{2}{c}{
Efficacy utility function - Equation - \ref{eq:pow_ind_util_eff} \vspace{0.2cm} }\\
$\overline{\pi}_E=0.5$ & Reference point where attitude to risk changes. $\pi_E \le 0.5$ is described as a loss and $\pi_E>0.5$ a gain\\ 
$\lambda_E=2$ & Loss aversion parameter, specified so that losses are twice as impactful as gains\\
$\alpha_{GE}=0.7$ & risk averse attitude to risk above the reference point (a gain)\\
$\alpha_{LE}=0.7$ & risk seeking attitude to risk below the reference point (a loss) \vspace{0.2cm}\\
\multicolumn{2}{c}{
Toxicity utility function - Equation \ref{eq:ind_util_tox} \vspace{0.2cm} }\\
$\overline{\pi}_T=0.35$ & Reference point where attitude to risk changes. $\pi_T \ge 0.35$ is described as a loss and $\pi_T<0.35$ a gain\\ 
$\lambda_T=2$ & Loss aversion parameter, specified so that losses are twice as impactful as gains\\
$\alpha_{GT}=0.7$ & risk averse attitude to risk below the reference point (a gain)  \\
$\alpha_{LT}=0.7$ & risk seeking attitude to risk above the reference point (a loss) \vspace{0.2cm} \\
\multicolumn{2}{c}{Joint utility function - Equation  \ref{eq:utili_indep} \vspace{0.2cm} }\\
$k_E=0.25$ & utility when $\pi_E=1$ and $\pi_T=1$\\
$k_T=0.15$ & utility when $\pi_E=0$ and $\pi_T=0$ \\
$(1-k_E-k_T)=0.6$ & positive interaction between marginal utility functions \\    
\bottomrule
\end{tabular}
\end{table}

The primary comparator\textit{EffToxU (2)} sets the utility of a patient experiencing both an efficacy and a toxicity event as $K(1,1)=0.25$ and the utility of experiencing neither as $K(0,0)=0.15$ (Appendix \ref{app:EffToxU}). Note that \textit{EffToxu} is a degenerate case of the R2DT design with $K(1,1)=k_E=0.25$, $K(0,0)=k_T=0.15$ and $u_E(\pi_E)=\pi_E$ and $u_T(\pi_T)=1-\pi_T$ i.e $\lambda_E=\lambda_T=\alpha_{GE}=\alpha_{LE}=\alpha_{GT}=\alpha_{LT}=1$, with  $\overline{\pi}_T$ and $\overline{\pi}_E$ becoming redundant as both marginal utility functions are linear (Appendix \ref{app:EffToxU} for proof). Stopping rules are applied as per \textit{R2DT (1)}. 

\subsubsection{R2DT Stopping Rule}
The effect of the R2DT stopping rule is explored by specifying the same utility function as the \textit{R2DT (1)} and adapting the stopping rule. \textit{R2DT (3)} applies the stopping criterion specified in Section \ref{sct:stopping_rules} as an admissibility rule labelled \textit{utility admissibility rule}. That is, at each decision point doses are excluded from choosing the maximum utility if there is insufficient evidence that a dose has acceptable levels of combined efficacy and toxicity. \textit{R2DT (4)} applies the utility trial stopping rule. This maximises the expected utility for all doses at each decision point with the trial stopping if all doses fail to satisfy the stopping rules given for \textit{R2DT (3)}. The threshold is $p_u=0.1$ for all designs according to Equation \ref{eq:R2DTstp}. Three contours have been specified and are listed in \ref{tab:Table_1}, and plotted in Appendix \ref{fig:scenerio_utility}. The admissibility rules accept lower utility for \textit{R2DT (3i)} and \textit{R2DT (3ii)} in comparison to the contour of \textit{R2DT (3iii)}, which declares higher utility unacceptable. Linking these differences to scenarios, some doses will be designated as acceptable according to one stopping rule while another stopping rule may say they are not acceptable.  

\subsubsection{Comparator Sensitivity}
There are two sensitivity analyses applied to the comparative \textit{EffToxU} method to demonstrate that conclusions are not just the result of a poorly specified comparator. The design with $K(1,1)=k_E=0.5$ and $K(0,0)=k_T=0.3$ has been stated as suitable in many settings \cite{Yuan2016a} and is specified in \textit{EffToxU (7)}. The ratio of$(k_E:k_T)$ is the same as \textit{EffToxU (2)} but the magnitude of $k_E+k_T=0.8$ is increased, suggesting a smaller interaction for $k_{ET}$ from Equation \ref{eq:utili_indep}.  \textit{EffTox (6)} applies the method of trade off contours \cite{Thall2004}, with specification of the design contour corresponding with the \textit{EffToxU (2)} equal utility contour passing through the reference point defined in \textit{R2DT (1)} and the points on the contour which have no toxicity and perfect efficacy.  

The stopping rule of R2DT is applied to the \textit{EffToxU (2)} comparator to give context. \textit{EffToxU (5)} explores the admissibility stopping rule in Equation \ref{eq:R2DTstp} using $u(0.5,0.35)= 0.42$ and $p_u=0.1$. Contour plots are provided in Appendix \ref{fig:scenerio_utility}.

\section{Results}
\subsection{Method Comparison}
The \textit{R2DT (1)} and \textit{EffToxU (2)} designs are simulated and contrasted in 10 scenarios (Table \ref{tab:Table_2}). To define which dose is the most desirable in any given scenario, doses are first excluded by the stopping rule, i.e any dose that has greater than $40\%$ toxicity or less than $50\%$ efficacy cannot be the optimum dose. The optimum dose is then defined by the maximum utility value. Scenarios 1 \& 2 have minimal toxicity and relatively steep efficacy with the $50mg/kg$ dose optimal. Both methods have very similar percentage of selection and numbers of patients treated at each dose. Scenarios 3 and 4 mirror the efficacy of scenarios 1 and 2 but increase the toxicity with both methods indicating the $40mg/kg$ dose as optimal. \textit{R2DT (1)} recommends the optimum dose more often particularly in scenario 3. 

Scenario 5 is steeply increasing with efficacy but is also very toxic, relative to the reference point, with the $20mg/kg$ dose optimal according to \textit{R2DT (1)} and the $50mg/kg$ dose according to \textit{EffToxU (2)}. In this scenario \textit{EffToxU (2)} has an equal utility with rounding to 2 decimal places at the $40mg/kg$ doseith. Without rounding the $50mg/kg$ has a marginally higher utility. In practice this means that in this scenario the doses are considered practically equivalent. When combining with the toxicity admissibility rule, however, all but the $20mg/kg$ for \textit{EffToxU(2)} are inadmissible. \textit{R2DT (1)} chooses the lower two doses more often under this scenario. Scenario 6 is flat for efficacy with the $20mg/kg$ dose optimal according to both utility functions. \textit{EffToxU (2)} out performs \textit{R2DT (1)} in terms of correct selection. Scenario 7 has an efficacy plateau at the $40mg/kg$ dose; \textit{R2DT (1)} strongly outperforms in this scenario. Scenario 8 has a steep increase in toxicity with the $30mg/kg$ optimal. \textit{R2DT (1)} out performs \textit{EffToxU (2)}. Scenario 9 is specified with all doses overly toxic. The two designs perform similarly despite \textit{EffToxU (2)} suggesting the $40mg/kg$ is optimal according to the utility function. The decision making process in this scenario is dominated by the admissibility rules which are identical between the designs. Scenario 10 is minimally efficacious for all doses with similar interpretation to the previous scenario.  

\begin{figure}[ht!]
\includegraphics[width=0.95\textwidth]{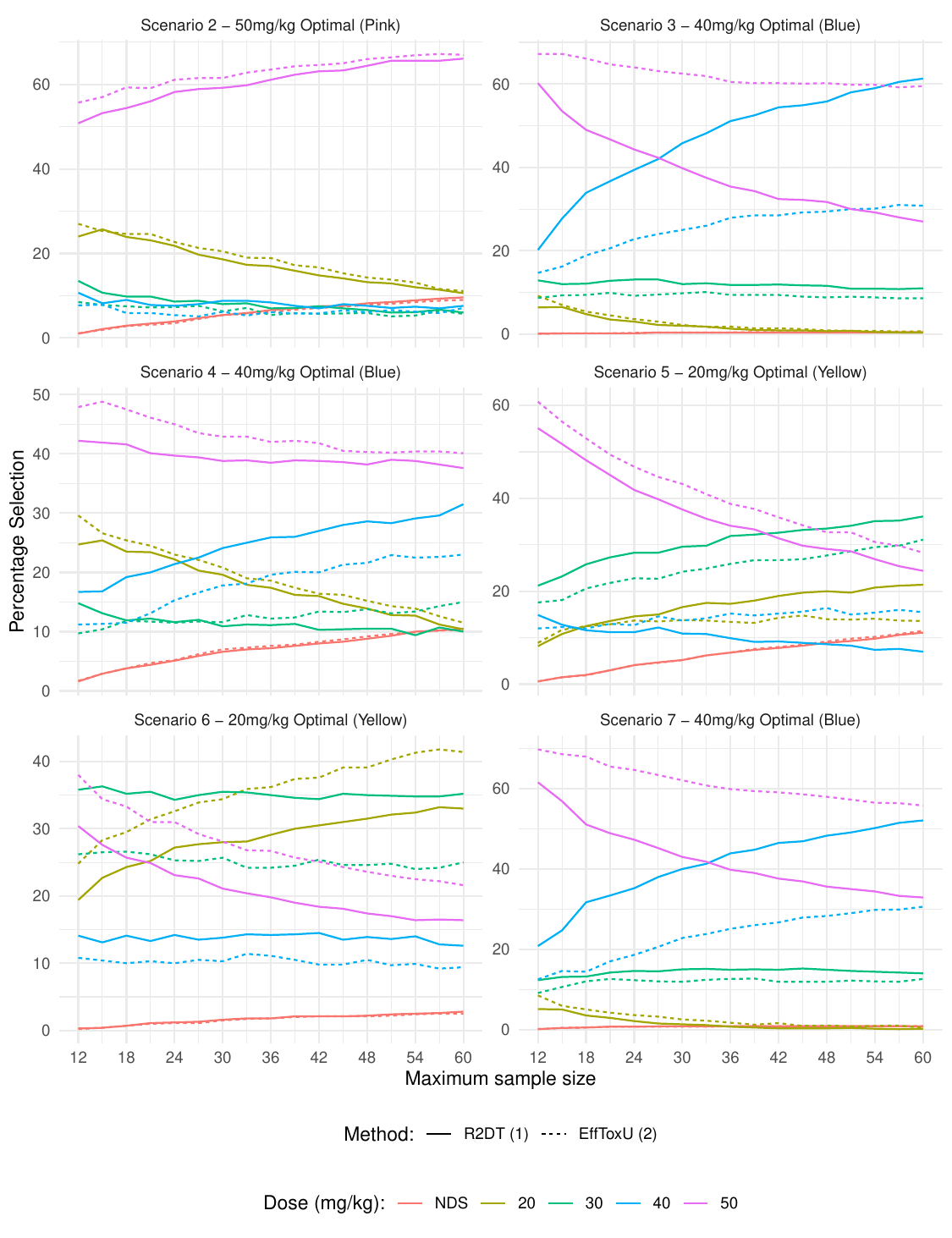}
  \caption[Simulation study results by sample size]{Percentage selection by sample size:
       Each scenario is plotted separately with solid lines representing \textit{R2DT (1)} method, dashed lines \textit{EfftoxU (2)} and the percentage of selection of each dose at each stage of the trial by colour.}
        \label{fig:sampsize}
    \end{figure}

The probability of selection of each dose for Scenarios 2 - 7 with the two methods is contrasted with sample size in Figure \ref{fig:sampsize}. In all scenarios after 12 patients the \textit{EffToxU (2)} has a greater proportion of simulated trials selecting the 50mg/kg dose as optimal suggesting that the \textit{R2DT (1)} method is initially more conservative in escalation. The probability of correct selection of both methods increase with sample size. The choice of 45 patients was deemed appropriate in this setting based upon the slower rate of improvement in accuracy after 45 patients and is seen as a clinically realistic sample size for this number of doses and setting.

\subsection{R2DT Stopping Rule}
The results of the simulation study inspecting the R2DT stopping rule are provided in Tables \ref{tab:R2DT_stop_1} and \ref{tab:R2DT_stop_2}. Specification of different stopping rules for R2DT makes minimal difference in scenarios 1, 3 and  7. In these scenarios some of the lower dose levels may be unacceptable but the main driver of design performance is the utility function, which is the same between each design. In general in the other scenarios the admissibility stopping rules (\textit{R2DT (3)}) are more likely to exclude doses and more likely to recommend stopping the trial in contrast to the trial stopping rule in (\textit{R2DT (4)}). This is a comparison between designs where the contour is the same, denoted by the same Roman numeral. The difference is slightly larger in designs requiring the highest utility, $(iii)$, but still less than $5\%$. This result is not unexpected as the only time that the decisions will differ is if a dose maximises the expected utility but also meets the threshold to be classed as inadmissible. In most instances the dose with the maximum expected utility will also be admissible. The two stopping rules will recommend stopping the trial at the same point given the same data however.     

In scenario 2 (\textit{R2DT (3iii)}) recommends stopping with no dose selected in $18\%$ of simulations. This is because the utility of $0.73$ for the $50mg/kg$ dose is close to the stopping contour with utility of $0.69$. In scenario 4 the trial is more likely to stop and select no dose with the R2DT stopping rules in contrast to \textit{R2DT (1)}. This is at the expense of selecting the $50mg/kg$ dose less frequently. There appears to be a noticeable difference for Scenario 5, where the stopping rule based upon utility for (\textit{R2DT (3i)}) suggests that all doses are acceptable (Utility at each dose is greater than the reference utility values) while the stopping rule based upon the individual probabilities would exclude all but the $20mg/kg$ dose. In scenarios 6, 8, 9 and 10 the contour stopping rules are more likely to end the trial without recommending a dose. This is proportional to how strict the stopping rule is with the designs needing a higher utility stopping more often. The rules are not directly comparable, and in practice which one is sensible would need clinical input for the given situation.

\subsection{Comparator Sensitivity}
\textit{EffToxU (7)} and \textit{EffTox (6)}, specified as a sensitivity for the specification of \textit{EffToxU (2)}, make little difference to the operating characteristics (Appendix Table \ref{tab:Table_3}). There is a difference in scenario 3 with \textit{EffToxU (7)} suggesting dose level 4 is optimal and selecting this dose level more often. 

Applying the R2DT stopping rule to EffTox in \textit{EffToxU (5)}, Tables \ref{tab:Table_5}, makes little difference to scenarios 1, 2, 3, 4, 6, 7 in comparison to \textit{EffToxU (2)}. In Scenario 5 the contour stipulates that all doses are acceptable and maximises more frequently to the $50mg/kg$ dose which has a toxicity of $51\%$. In scenarios 8, 9 and 10 the \textit{EffToxU (5)} suggest that higher doses have acceptable toxicity given high efficacy. Take scenario 9, for example. It is only the $20mg/kg$ dose that has unacceptable utility in contrast to all doses in \textit{EffToxU (2)}. This results in the utility stopping rule more frequently recommending higher doses and a lower proportion of trials stopping early without selecting a dose. Similarly, in scenario 10 the alternative stopping rule recommends the $50mg/kg$ a high proportion of the time. Here the $50mg/kg$ is acceptable according to the specified stopping rule. It is unlikely that a contour could be specified that accommodates a threshold for unacceptability. This set of simulations highlights the dependence of the EffTox method on the stopping rules to restrict treating at doses with unacceptable toxicity and/or efficacy.  
\newpage
\begin{table}[ht!]

\caption[Simulation study results comparison between RT2D and EffToxU]{\label{tab:Table_2}Comparison between RT2D and EffToxU: data of form: [utility at scenario probability $(\pi_E, \pi_T)$] percentage selection (average number of patients treated). Percentage of trials with no dose selected abbreviated to NDS. Bold indicates optimal dose or whether trial
should recommend not selecting a dose (and stop early)}
\centering
\begin{tabular}[t]{llllll}
\toprule
\multicolumn{1}{c}{ } & \multicolumn{4}{c}{Dose (mg/kg)} & \multicolumn{1}{c}{ } \\
\cmidrule(l{3pt}r{3pt}){2-5}
\multicolumn{1}{l}{Method} & \multicolumn{1}{c}{20} & \multicolumn{1}{c}{30} & \multicolumn{1}{c}{40} & \multicolumn{1}{c}{50} & \multicolumn{1}{c}{NDS}\\
\midrule
\addlinespace[0.3em]
\multicolumn{6}{c}{\textbf{Scenario 1} $(\pi_E, \pi_T)$}\\
\multicolumn{1}{c}{\hspace{1em}} & \multicolumn{1}{c}{(0.3, 0.05)} & \multicolumn{1}{c}{(0.57, 0.08)} & \multicolumn{1}{c}{(0.75, 0.12)} & \multicolumn{1}{c}{\textbf{(0.85, 0.15)}} & \multicolumn{1}{c}{}\\
\midrule
\hspace{1em}R2DT (1) & {}[0.41]  0.9 (5.1) & {}[0.76] 4 (4.9) & {}[0.85] 8.6 (5.8) & {}[0.88] 86.2 (29.1) & 0.3\\
\hspace{1em}EffToxU (2) & {}[0.39]  1.5 (4.8) & {}[0.60] 4.1 (5.3) & {}[0.72] 3.8 (4.1) & {}[0.77] 90.4 (30.8) & 0.2\\
\addlinespace[0.3em]
\multicolumn{6}{c}{\textbf{Scenario 2} $(\pi_E, \pi_T)$}\\
\multicolumn{1}{c}{\hspace{1em}} & \multicolumn{1}{c}{(0.37, 0.05)} & \multicolumn{1}{c}{(0.45, 0.08)} & \multicolumn{1}{c}{(0.51, 0.12)} & \multicolumn{1}{c}{\textbf{(0.55, 0.15)}} & \multicolumn{1}{c}{}\\
\midrule
\hspace{1em}R2DT (1) & {}[0.49] 14.1 (11.6) & {}[0.58] 7 (6.2) & {}[0.70] 8 (5.1) & {}[0.73] 63.3 (20.6) & 7.6\\
\hspace{1em}EffToxU (2) & {}[0.45] 15.3 (11.6) & {}[0.50] 5.8 (5.9) & {}[0.53] 6.5 (4.2) & {}[0.55] 65 (21.8) & 7.4\\
\addlinespace[0.3em]
\multicolumn{6}{c}{\textbf{Scenario 3} $(\pi_E, \pi_T)$}\\
\multicolumn{1}{c}{\hspace{1em}} & \multicolumn{1}{c}{(0.3, 0.05)} & \multicolumn{1}{c}{(0.57, 0.13)} & \multicolumn{1}{c}{\textbf{(0.75, 0.23)}} & \multicolumn{1}{c}{(0.85, 0.35)} & \multicolumn{1}{c}{}\\
\midrule
\hspace{1em}R2DT (1) & {}[0.41] 0.9 (4.9) & {}[0.75] 11.7 (7.4) & {}[0.80] 54.9 (16.4) & {}[0.76] 32.2 (16.2) & 0.4\\
\hspace{1em}EffToxU (2) & {}[0.39] 1.2 (4.8) & {}[0.57] 9 (6.7) & {}[0.65] 29.2 (10.2) & {}[0.64] 60.1 (23.1) & 0.5\\
\addlinespace[0.3em]
\multicolumn{6}{c}{\textbf{Scenario 4} $(\pi_E, \pi_T)$}\\
\multicolumn{1}{c}{\hspace{1em}} & \multicolumn{1}{c}{(0.37, 0.05)} & \multicolumn{1}{c}{(0.45, 0.13)} & \multicolumn{1}{c}{\textbf{(0.51, 0.23)}} & \multicolumn{1}{c}{(0.55, 0.35)} & \multicolumn{1}{c}{}\\
\midrule
\hspace{1em}R2DT (1) & {}[0.49] 14.7 (11.7) & {}[0.57] 10.4 (7.1) & {}[0.66] 28 (9.7) & {}[0.63] 38.6 (14.6) & 8.3\\
\hspace{1em}EffToxU (2) & {}[0.45] 16.2 (11.8) & {}[0.48] 13.3 (7.5) & {}[0.48] 21.3 (7.4) & {}[0.45] 40.5 (16.4) & 8.7\\
\addlinespace[0.3em]
\multicolumn{6}{c}{\textbf{Scenario 5} $(\pi_E, \pi_T)$}\\
\multicolumn{1}{c}{\hspace{1em}} & \multicolumn{1}{c}{\textbf{(0.55, 0.35)}} & \multicolumn{1}{c}{(0.75, 0.42)} & \multicolumn{1}{c}{(0.85, 0.47)} & \multicolumn{1}{c}{(0.9, 0.51)} & \multicolumn{1}{c}{}\\
\midrule
\hspace{1em}R2DT (1) & {}[0.63] 19.7 (8.5) & {}[0.62] 33.2 (13.2) & {}[0.60] 8.9 (6.5) & {}[0.58] 29.8 (15.3) & 8.3\\
\hspace{1em}EffToxU (2) & {}[0.45] 14.8 (7.6) & {}[0.54] 26.9 (11.4) & {}[0.56] 15.6 (7.3) & {}[0.56] 34.2 (17.2) & 8.5\\
\addlinespace[0.3em]
\multicolumn{6}{c}{\textbf{Scenario 6} $(\pi_E, \pi_T)$}\\
\multicolumn{1}{c}{\hspace{1em}} & \multicolumn{1}{c}{\textbf{(0.6, 0.26)}} & \multicolumn{1}{c}{(0.62, 0.35)} & \multicolumn{1}{c}{(0.63, 0.42)} & \multicolumn{1}{c}{(0.64, 0.48)} & \multicolumn{1}{c}{}\\
\midrule
\hspace{1em}R2DT (1) & {}[0.72] 31 (13.1) & {}[0.67] 35.2 (16) & {}[0.57] 13.5 (6.8) & {}[0.52] 18.1 (8.6) & 2.1\\
\hspace{1em}EffToxU (2) & {}[0.53] 39.1 (14.8) & {}[0.49] 24.6 (12.8) & {}[0.46] 9.8 (5.8) & {}[0.44] 24.3 (11.2) & 2.1\\
\addlinespace[0.3em]
\multicolumn{6}{c}{\textbf{Scenario 7} $(\pi_E, \pi_T)$}\\
\multicolumn{1}{c}{\hspace{1em}} & \multicolumn{1}{c}{(0.26, 0.05)} & \multicolumn{1}{c}{(0.6, 0.13)} & \multicolumn{1}{c}{\textbf{(0.7, 0.23)}} & \multicolumn{1}{c}{(0.7, 0.35)} & \multicolumn{1}{c}{}\\
\midrule
\hspace{1em}R2DT (1) & {}[0.37] 0.3 (4.4) & {}[0.77] 15.2 (8) & {}[0.78] 46.9 (14.9) & {}[0.70] 36.9 (17.4) & 0.8\\
\hspace{1em}EffToxU (2) & {}[0.36] 0.9 (4.6) & {}[0.59] 11.9 (7.4) & {}[0.61] 27.9 (9.4) & {}[0.54] 58.6 (23.4) & 0.8\\
\addlinespace[0.3em]
\multicolumn{6}{c}{\textbf{Scenario 8} $(\pi_E, \pi_T)$}\\
\multicolumn{1}{c}{\hspace{1em}} & \multicolumn{1}{c}{(0.26, 0.18)} & \multicolumn{1}{c}{\textbf{(0.6, 0.35)}} & \multicolumn{1}{c}{(0.7, 0.5)} & \multicolumn{1}{c}{(0.7, 0.62)} & \multicolumn{1}{c}{}\\
\midrule
\hspace{1em}R2DT (1) & {}[0.35] 3.4 (5.6) & {}[0.66] 61.4 (18.2) & {}[0.53] 22.2 (11) & {}[0.44] 6.2 (8.8) & 6.8\\
\hspace{1em}EffToxU (2) & {}[0.32] 3.9 (6.3) & {}[0.48] 50.8 (14.4) & {}[0.46] 26.5 (10.6) & {}[0.39] 11.8 (12.3) & 7\\
\addlinespace[0.3em]
\multicolumn{6}{c}{\textbf{Scenario 9} $(\pi_E, \pi_T)$}\\
\multicolumn{1}{c}{\hspace{1em}} & \multicolumn{1}{c}{(0.55, 0.45)} & \multicolumn{1}{c}{(0.75, 0.57)} & \multicolumn{1}{c}{(0.85, 0.64)} & \multicolumn{1}{c}{(0.9, 0.7)} & \multicolumn{1}{c}{}\\
\midrule
\hspace{1em}R2DT (1) & {}[0.51] 32.4 (13.1) & {}[0.49] 10.8 (8.5) & {}[0.46] 0.9 (4) & {}[0.43] 2.9 (8.3) & \textbf{52.9}\\
\hspace{1em}EffToxU (2) & {}[0.40] 29.4 (12.3) & {}[0.45] 13.2 (8.9) & {}[0.45] 2 (4.3) & {}[0.43] 2.9 (8.6) & \textbf{52.5}\\
\addlinespace[0.3em]
\multicolumn{6}{c}{\textbf{Scenario 10} $(\pi_E, \pi_T)$}\\
\multicolumn{1}{c}{\hspace{1em}} & \multicolumn{1}{c}{(0.2, 0.05)} & \multicolumn{1}{c}{(0.3, 0.08)} & \multicolumn{1}{c}{(0.38, 0.12)} & \multicolumn{1}{c}{(0.45, 0.15)} & \multicolumn{1}{c}{}\\
\midrule
\hspace{1em}R2DT (1) & {}[0.31] 1.6 (6.2) & {}[0.40] 0.6 (3.7) & {}[0.48] 0.9 (3.7) & {}[0.57] 51.1 (21.3) & \textbf{45.7}\\
\hspace{1em}EffToxU (2) & {}[0.31] 1.5 (6.1) & {}[0.38] 0.9 (3.7) & {}[0.43] 1.5 (3.6) & {}[0.47] 51.1 (21.7) & \textbf{44.9}\\
\bottomrule
\end{tabular}
\end{table}

\newpage
\clearpage
\begin{table}[ht!]
\caption[Simulation results evaluating stopping rules of R2DT(scenarios 1:5)]{\label{tab:R2DT_stop_1}R2DT Stopping rule: data of form: [utility at scenario probability $(\pi_E, \pi_T)$] percentage selection (average number of patients treated). Percentage of trials with no dose selected abbreviated to NDS. Bold indicates optimal dose or whether trial should recommend not selecting a dose (and stop early)}
\centering
\begin{tabular}[t]{llllll}
\toprule
\multicolumn{1}{c}{ } & \multicolumn{4}{c}{Dose (mg/kg)} & \multicolumn{1}{c}{ } \\
\cmidrule(l{3pt}r{3pt}){2-5}
\multicolumn{1}{l}{Method} & \multicolumn{1}{c}{20} & \multicolumn{1}{c}{30} & \multicolumn{1}{c}{40} & \multicolumn{1}{c}{50} & \multicolumn{1}{c}{NDS}\\
\midrule
\addlinespace[0.3em]
\multicolumn{6}{c}{\textbf{Scenario 1} $(\pi_E, \pi_T)$}\\
\multicolumn{1}{c}{\hspace{1em}} & \multicolumn{1}{c}{(0.3, 0.05)} & \multicolumn{1}{c}{(0.57, 0.08)} & \multicolumn{1}{c}{(0.75, 0.12)} & \multicolumn{1}{c}{(0.85, 0.15)} & \multicolumn{1}{c}{}\\
\midrule
\hspace{1em}R2DT (1) & {}[0.41] 0.9 (5.3) & {}[0.76] 4.1 (5) & {}[0.85] 8.7 (5.8) & {}[\textbf{0.88}] 86.2 (29) & 0.1\\
\hspace{1em}R2DT (3i) & {}[0.41] 1.7 (5.5) & {}[0.76] 4.3 (5) & {}[0.85] 8.1 (5.8) & {}[\textbf{0.88}] 85.7 (28.6) & 0.3\\
\hspace{1em}R2DT (3ii) & {}[0.41] 1.5 (5.3) & {}[0.76] 4 (5) & {}[0.85] 8.2 (5.8) & {}[\textbf{0.88}] 85.9 (28.8) & 0.4\\
\hspace{1em}R2DT (3iii) & {}[0.41] 0.9 (5.1) & {}[0.76] 4 (4.9) & {}[0.85] 8.1 (5.8) & {}[\textbf{0.88}] 83.9 (28.1) & 3\\
\hspace{1em}R2DT (4i) & {}[0.41] 3.4 (5.8) & {}[0.76] 4 (4.9) & {}[0.85] 8 (5.8) & {}[\textbf{0.88}] 84.4 (28.5) & 0.2\\
\hspace{1em}R2DT (4ii) & {}[0.41] 3.2 (5.7) & {}[0.76] 4 (4.9) & {}[0.85] 8 (5.8) & {}[\textbf{0.88}] 84.4 (28.4) & 0.4\\
\hspace{1em}R2DT (4iii) & {}[0.41] 3.2 (5.7) & {}[0.76] 4 (4.9) & {}[0.85] 8 (5.7) & {}[\textbf{0.88}] 82 (27.6) & 2.9\\
\addlinespace[0.3em]
\multicolumn{6}{c}{\textbf{Scenario 2} $(\pi_E, \pi_T)$}\\
\multicolumn{1}{c}{\hspace{1em}} & \multicolumn{1}{c}{(0.37, 0.05)} & \multicolumn{1}{c}{(0.45, 0.08)} & \multicolumn{1}{c}{(0.51, 0.12)} & \multicolumn{1}{c}{(0.55, 0.15)} & \multicolumn{1}{c}{}\\
\midrule
\hspace{1em}R2DT (1) & {}[0.49] 16 (12.2) & {}[0.58] 7.2 (6.3) & {}[0.70] 7.8 (5) & {}[\textbf{0.73}] 64.5 (20.6) & 4.5\\
\hspace{1em}R2DT (3i) & {}[0.49] 18 (12.6) & {}[0.58] 7 (6.2) & {}[0.70] 7 (5) & {}[\textbf{0.73}] 62.6 (20.1) & 5.3\\
\hspace{1em}R2DT (3ii) & {}[0.49] 15.3 (11.8) & {}[0.58] 6.7 (6.1) & {}[0.70] 7 (5) & {}[\textbf{0.73}] 62.4 (20.1) & 8.6\\
\hspace{1em}R2DT (3iii) & {}[0.49] 11.7 (10.8) & {}[0.58] 6.6 (6) & {}[0.70] 7.2 (4.9) & {}[\textbf{0.73}] 56.5 (18.6) & 18\\
\hspace{1em}R2DT (4i) & {}[0.49] 21.4 (13.1) & {}[0.58] 6.8 (6.1) & {}[0.70] 6.8 (4.9) & {}[\textbf{0.73}] 60.7 (19.8) & 4.4\\
\hspace{1em}R2DT (4ii) & {}[0.49] 19.9 (12.8) & {}[0.58] 6.7 (6.1) & {}[0.70] 6.7 (4.9) & {}[\textbf{0.73}] 59.4 (19.5) & 7.3\\
\hspace{1em}R2DT (4iii) & {}[0.49] 18.1 (12.4) & {}[0.58] 6.6 (6.1) & {}[0.70] 6.7 (4.7) & {}[\textbf{0.73}] 53 (17.6) & 15.6\\
\addlinespace[0.3em]
\multicolumn{6}{c}{\textbf{Scenario 3} $(\pi_E, \pi_T)$}\\
\multicolumn{1}{c}{\hspace{1em}} & \multicolumn{1}{c}{(0.3, 0.05)} & \multicolumn{1}{c}{(0.57, 0.13)} & \multicolumn{1}{c}{(0.75, 0.23)} & \multicolumn{1}{c}{(0.85, 0.35)} & \multicolumn{1}{c}{}\\
\midrule
\hspace{1em}R2DT (1) & {}[0.41] 0.9 (5) & {}[0.75] 11.7 (7.4) & {}[\textbf{0.80}] 54.9 (16.3) & {}[0.76] 32.4 (16.2) & 0.1\\
\hspace{1em}R2DT (3i) & {}[0.41] 1 (5.2) & {}[0.75] 12.5 (7.4) & {}[\textbf{0.80}] 54.9 (16.3) & {}[0.76] 30.9 (16) & 0.7\\
\hspace{1em}R2DT (3ii) & {}[0.41] 0.9 (5) & {}[0.75] 12 (7.4) & {}[\textbf{0.80}] 55.2 (16.3) & {}[0.76] 30.6 (15.9) & 1.2\\
\hspace{1em}R2DT (3iii) & {}[0.41] 0.6 (4.7) & {}[0.75] 11.8 (7.4) & {}[\textbf{0.80}] 53.9 (16) & {}[0.76] 28.5 (15.2) & 5.1\\
\hspace{1em}R2DT (4i) & {}[0.41] 2.5 (5.4) & {}[0.75] 12 (7.4) & {}[\textbf{0.80}] 54.1 (16.1) & {}[0.76] 30.6 (15.9) & 0.7\\
\hspace{1em}R2DT (4ii) & {}[0.41] 2.4 (5.4) & {}[0.75] 12 (7.4) & {}[\textbf{0.80}] 53.9 (16.1) & {}[0.76] 30.4 (15.8) & 1.2\\
\hspace{1em}R2DT (4iii) & {}[0.41] 2.1 (5.3) & {}[0.75] 11.9 (7.4) & {}[\textbf{0.80}] 52.6 (15.8) & {}[0.76] 28.4 (14.9) & 4.9\\
\addlinespace[0.3em]
\multicolumn{6}{c}{\textbf{Scenario 4} $(\pi_E, \pi_T)$}\\
\multicolumn{1}{c}{\hspace{1em}} & \multicolumn{1}{c}{(0.37, 0.05)} & \multicolumn{1}{c}{(0.45, 0.13)} & \multicolumn{1}{c}{(0.51, 0.23)} & \multicolumn{1}{c}{(0.55, 0.35)} & \multicolumn{1}{c}{}\\
\midrule
\hspace{1em}R2DT (1) & {}[0.49] 17.8 (12.6) & {}[0.57] 10.9 (7.3) & {}[\textbf{0.66}] 27.4 (9.4) & {}[0.63] 39 (14.8) & 5\\
\hspace{1em}R2DT (3i) & {}[0.49] 20.2 (12.8) & {}[0.57] 9.7 (7) & {}[\textbf{0.66}] 25.1 (9) & {}[0.63] 34.6 (13.9) & 10.4\\
\hspace{1em}R2DT (3ii) & {}[0.49] 17.4 (12) & {}[0.57] 9.2 (6.9) & {}[\textbf{0.66}] 24.6 (9) & {}[0.63] 33.1 (13.6) & 15.8\\
\hspace{1em}R2DT (3iii) & {}[0.49] 14 (10.9) & {}[0.57] 8.7 (6.8) & {}[0.66] 21.4 (8.4) & {}[0.63] 25.7 (11.9) & \textbf{30.2}\\
\hspace{1em}R2DT (4i) & {}[0.49] 24.5 (13.6) & {}[0.57] 9.4 (7) & {}[\textbf{0.66}] 23.8 (8.8) & {}[0.63] 32.8 (13.4) & 9.5\\
\hspace{1em}R2DT (4ii) & {}[0.49] 23.3 (13.3) & {}[0.57] 9.2 (6.9) & {}[\textbf{0.66}] 23 (8.7) & {}[0.63] 31 (13) & 13.6\\
\hspace{1em}R2DT (4iii) & {}[0.49] 20.8 (12.8) & {}[0.57] 8.8 (6.8) & {}[0.66] 21.2 (8.2) & {}[0.63] 23.8 (11) & \textbf{25.5}\\
\addlinespace[0.3em]
\multicolumn{6}{c}{\textbf{Scenario 5} $(\pi_E, \pi_T)$}\\
\multicolumn{1}{c}{\hspace{1em}} & \multicolumn{1}{c}{(0.55, 0.35)} & \multicolumn{1}{c}{(0.75, 0.42)} & \multicolumn{1}{c}{(0.85, 0.47)} & \multicolumn{1}{c}{(0.9, 0.51)} & \multicolumn{1}{c}{}\\
\midrule
\hspace{1em}R2DT (1) & {}[\textbf{0.63}] 19.2 (8.4) & {}[0.62] 33.8 (13.2) & {}[0.60] 9.1 (6.6) & {}[0.58] 33.1 (16) & 4.8\\
\hspace{1em}R2DT (3i) & {}[\textbf{0.63}] 15.3 (7.8) & {}[0.62] 29 (12.4) & {}[0.60] 9.6 (6.8) & {}[0.58] 40.6 (16.9) & 5.6\\
\hspace{1em}R2DT (3ii) & {}[\textbf{0.63}] 15.2 (7.7) & {}[0.62] 28.4 (12.1) & {}[0.60] 9 (6.7) & {}[0.58] 36.2 (16.2) & 11.1\\
\hspace{1em}R2DT (3iii) & {}[0.63] 13.2 (7.4) & {}[0.62] 23.6 (11.3) & {}[0.60] 7.7 (6.4) & {}[0.58] 26.1 (14.3) & \textbf{29.4}\\
\hspace{1em}R2DT (4i) & {}[\textbf{0.63}] 14.2 (7.5) & {}[0.62] 30.1 (12.6) & {}[0.60] 9.8 (6.8) & {}[0.58] 40.7 (16.9) & 5.2\\
\hspace{1em}R2DT (4ii) & {}[\textbf{0.63}] 13.9 (7.5) & {}[0.62] 29.5 (12.5) & {}[0.60] 9.4 (6.8) & {}[0.58] 36.9 (16.1) & 10.2\\
\hspace{1em}R2DT (4iii) & {}[0.63] 11.7 (7) & {}[0.62] 26 (11.9) & {}[0.60] 8.4 (6.5) & {}[0.58] 27.5 (14.3) & \textbf{26.4}\\
\bottomrule
\end{tabular}
\end{table}

\newpage
\begin{table}[ht!]
\caption[Simulation results evaluating novel stopping rules of R2DT (scenarios 6:10)]{\label{tab:R2DT_stop_2}R2DT stopping rule continued: data of form: [utility at scenario probability $(\pi_E, \pi_T)$] percentage selection (average number of patients treated). Percentage of trials with no dose selected abbreviated to NDS. Bold indicates optimal dose or whether trial should recommend not selecting a dose (and stop early)}
\centering
\begin{tabular}[t]{llllll}
\toprule
\multicolumn{1}{c}{ } & \multicolumn{4}{c}{Dose (mg/kg)} & \multicolumn{1}{c}{ } \\
\cmidrule(l{3pt}r{3pt}){2-5}
\multicolumn{1}{l}{Method} & \multicolumn{1}{c}{20} & \multicolumn{1}{c}{30} & \multicolumn{1}{c}{40} & \multicolumn{1}{c}{50} & \multicolumn{1}{c}{NDS}\\
\midrule
\addlinespace[0.3em]
\multicolumn{6}{c}{\textbf{Scenario 6} $(\pi_E, \pi_T)$}\\
\multicolumn{1}{c}{\hspace{1em}} & \multicolumn{1}{c}{(0.6, 0.26)} & \multicolumn{1}{c}{(0.62, 0.35)} & \multicolumn{1}{c}{(0.63, 0.42)} & \multicolumn{1}{c}{(0.64, 0.48)} & \multicolumn{1}{c}{}\\
\midrule
\hspace{1em}R2DT (1) & {}[\textbf{0.72}] 30.6 (13.2) & {}[0.67] 35.1 (16) & {}[0.57] 13.8 (6.8) & {}[0.52] 19.4 (8.9) & 1\\
\hspace{1em}R2DT (3i) & {}[\textbf{0.72}] 30 (13) & {}[0.67] 33.1 (15.6) & {}[0.57] 12.3 (6.5) & {}[0.52] 17.8 (8.4) & 6.7\\
\hspace{1em}R2DT (3ii) & {}[\textbf{0.72}] 30 (12.8) & {}[0.67] 32.3 (15.3) & {}[0.57] 11 (6.3) & {}[0.52] 15.7 (8) & 11.1\\
\hspace{1em}R2DT (3iii) & {}[\textbf{0.72}] 27.6 (12.1) & {}[0.67] 27.8 (14.3) & {}[0.57] 9.2 (6) & {}[0.52] 11.3 (7.1) & 24.1\\
\hspace{1em}R2DT (4i) & {}[\textbf{0.72}] 29.3 (12.9) & {}[0.67] 32.9 (15.6) & {}[0.57] 12.4 (6.5) & {}[0.52] 18.6 (8.4) & 6.8\\
\hspace{1em}R2DT (4ii) & {}[\textbf{0.72}] 28.1 (12.5) & {}[0.67] 32.7 (15.6) & {}[0.57] 12.1 (6.5) & {}[0.52] 16.4 (7.9) & 10.6\\
\hspace{1em}R2DT (4iii) & {}[\textbf{0.72}] 24.6 (11.6) & {}[0.67] 31.6 (15.3) & {}[0.57] 10.5 (6.1) & {}[0.52] 10.9 (6.7) & 22.4\\
\addlinespace[0.3em]
\multicolumn{6}{c}{\textbf{Scenario 7} $(\pi_E, \pi_T)$}\\
\multicolumn{1}{c}{\hspace{1em}} & \multicolumn{1}{c}{(0.26, 0.05)} & \multicolumn{1}{c}{(0.6, 0.13)} & \multicolumn{1}{c}{(0.7, 0.23)} & \multicolumn{1}{c}{(0.7, 0.35)} & \multicolumn{1}{c}{}\\
\midrule
\hspace{1em}R2DT (1) & {}[0.37] 0.4 (4.5) & {}[0.77] 15 (8) & {}[\textbf{0.78}] 46.8 (14.9) & {}[0.70] 37.4 (17.5) & 0.3\\
\hspace{1em}R2DT (3i) & {}[0.37] 0.6 (4.6) & {}[0.77] 14.4 (7.8) & {}[\textbf{0.78}] 46.2 (14.9) & {}[0.70] 37.5 (17.3) & 1.4\\
\hspace{1em}R2DT (3ii) & {}[0.37] 0.5 (4.5) & {}[0.77] 14.2 (7.8) & {}[\textbf{0.78}] 46.1 (14.9) & {}[0.70] 36.8 (17.2) & 2.4\\
\hspace{1em}R2DT (3iii) & {}[0.37] 0.5 (4.4) & {}[0.77] 14.5 (7.8) & {}[\textbf{0.78}] 44.9 (14.7) & {}[0.70] 33.2 (16.1) & 6.9\\
\hspace{1em}R2DT (4i) & {}[0.37] 1.3 (4.8) & {}[0.77] 14.4 (7.8) & {}[\textbf{0.78}] 45.9 (14.8) & {}[0.70] 37.1 (17.2) & 1.4\\
\hspace{1em}R2DT (4ii) & {}[0.37] 1.1 (4.8) & {}[0.77] 14.3 (7.8) & {}[\textbf{0.78}] 45.8 (14.8) & {}[0.70] 36.6 (17.1) & 2.2\\
\hspace{1em}R2DT (4iii) & {}[0.37] 1.1 (4.7) & {}[0.77] 14.2 (7.7) & {}[\textbf{0.78}] 44.9 (14.5) & {}[0.70] 32.8 (15.9) & 7\\
\addlinespace[0.3em]
\multicolumn{6}{c}{\textbf{Scenario 8} $(\pi_E, \pi_T)$}\\
\multicolumn{1}{c}{\hspace{1em}} & \multicolumn{1}{c}{(0.26, 0.18)} & \multicolumn{1}{c}{(0.6, 0.35)} & \multicolumn{1}{c}{(0.7, 0.5)} & \multicolumn{1}{c}{(0.7, 0.62)} & \multicolumn{1}{c}{}\\
\midrule
\hspace{1em}R2DT (1) & {}[0.35] 3.7 (5.7) & {}[\textbf{0.66}] 59.4 (17.8) & {}[0.53] 25 (11.3) & {}[0.44] 7.9 (9.5) & 4.1\\
\hspace{1em}R2DT (3i) & {}[0.35] 2.5 (5.3) & {}[\textbf{0.66}] 43.8 (15.4) & {}[0.53] 18.6 (10) & {}[0.44] 7.6 (8.6) & 27.6\\
\hspace{1em}R2DT (3ii) & {}[0.35] 1.9 (5.3) & {}[\textbf{0.66}] 41.3 (14.9) & {}[0.53] 14.1 (9.1) & {}[0.44] 4.4 (7.5) & 38.2\\
\hspace{1em}R2DT (3iii) & {}[0.35] 2.1 (5) & {}[0.66] 28.3 (12.4) & {}[0.53] 7.6 (7.6) & {}[0.44] 2.5 (6.1) & \textbf{59.7}\\
\hspace{1em}R2DT (4i) & {}[0.35] 2.8 (5.3) & {}[\textbf{0.66}] 41.3 (15.1) & {}[0.53] 21.9 (10.5) & {}[0.44] 9.2 (8.8) & 24.8\\
\hspace{1em}R2DT (4ii) & {}[0.35] 2.2 (5.2) & {}[\textbf{0.66}] 39.4 (14.7) & {}[0.53] 18 (9.8) & {}[0.44] 5.7 (7.7) & 34.8\\
\hspace{1em}R2DT (4iii) & {}[0.35] 1.2 (4.8) & {}[0.66] 32 (13.2) & {}[0.53] 10.8 (8.1) & {}[0.44] 2 (5.9) & \textbf{54}\\
\addlinespace[0.3em]
\multicolumn{6}{c}{\textbf{Scenario 9} $(\pi_E, \pi_T)$}\\
\multicolumn{1}{c}{\hspace{1em}} & \multicolumn{1}{c}{(0.55, 0.45)} & \multicolumn{1}{c}{(0.75, 0.57)} & \multicolumn{1}{c}{(0.85, 0.64)} & \multicolumn{1}{c}{(0.9, 0.7)} & \multicolumn{1}{c}{}\\
\midrule
\hspace{1em}R2DT (1) & {}[0.51] 39.6 (14.2) & {}[0.49] 13.6 (9.1) & {}[0.46] 1.3 (4.1) & {}[0.43] 4.2 (9.7) & \textbf{41.2}\\
\hspace{1em}R2DT (3i) & {}[0.51] 21.1 (10.4) & {}[0.49] 12.6 (8.4) & {}[0.46] 1.8 (4.3) & {}[0.43] 10.3 (11.5) & \textbf{54.2}\\
\hspace{1em}R2DT (3ii) & {}[0.51] 16.7 (9.6) & {}[0.49] 8.6 (7.4) & {}[0.46] 1.1 (4.1) & {}[0.43] 5.8 (9.5) & \textbf{67.8}\\
\hspace{1em}R2DT (3iii) & {}[0.51] 8.6 (7.8) & {}[0.49] 3.4 (6) & {}[0.46] 0.7 (3.8) & {}[0.43] 1.8 (7.2) & \textbf{85.6}\\
\hspace{1em}R2DT (4i) & {}[0.51] 17.2 (9.5) & {}[0.49] 18.2 (9.4) & {}[0.46] 2.5 (4.4) & {}[0.43] 13.6 (11.9) & \textbf{48.5}\\
\hspace{1em}R2DT (4ii) & {}[0.51] 13.8 (8.6) & {}[0.49] 14.3 (8.8) & {}[0.46] 1.8 (4.2) & {}[0.43] 7.6 (9.9) & \textbf{62.5}\\
\hspace{1em}R2DT (4iii) & {}[0.51] 7.8 (7.2) & {}[0.49] 7.8 (7.3) & {}[0.46] 0.7 (3.8) & {}[0.43] 2.5 (7.3) & \textbf{81.2}\\
\addlinespace[0.3em]
\multicolumn{6}{c}{\textbf{Scenario 10} $(\pi_E, \pi_T)$}\\
\multicolumn{1}{c}{\hspace{1em}} & \multicolumn{1}{c}{(0.2, 0.05)} & \multicolumn{1}{c}{(0.3, 0.08)} & \multicolumn{1}{c}{(0.38, 0.12)} & \multicolumn{1}{c}{(0.45, 0.15)} & \multicolumn{1}{c}{}\\
\midrule
\hspace{1em}R2DT (1) & {}[0.31] 2.3 (6.6) & {}[0.40] 1 (3.8) & {}[0.48] 0.9 (3.7) & {}[0.57] 60.5 (23.3) & \textbf{35.3}\\
\hspace{1em}R2DT (3i) & {}[0.31] 2 (6.6) & {}[0.40] 0.5 (3.7) & {}[0.48] 1 (3.6) & {}[0.57] 59.4 (22.4) & \textbf{37.2}\\
\hspace{1em}R2DT (3ii) & {}[0.31] 1.3 (6.2) & {}[0.40] 0.6 (3.6) & {}[0.48] 0.9 (3.6) & {}[0.57] 50.2 (20.5) & \textbf{46.9}\\
\hspace{1em}R2DT (3iii) & {}[0.31] 1.2 (5.5) & {}[0.40] 0.2 (3.6) & {}[0.48] 0.9 (3) & {}[0.57] 32.7 (14.8) & \textbf{65}\\
\hspace{1em}R2DT (4i) & {}[0.31] 2.9 (7.1) & {}[0.40] 0.6 (3.6) & {}[0.48] 1.2 (3.6) & {}[0.57] 60.5 (22.4) & \textbf{34.8}\\
\hspace{1em}R2DT (4ii) & {}[0.31] 2.4 (6.8) & {}[0.40] 0.5 (3.6) & {}[0.48] 1.2 (3.6) & {}[0.57] 51.6 (20.3) & \textbf{44.2}\\
\hspace{1em}R2DT (4iii) & {}[0.31] 2.1 (6.5) & {}[0.40] 0.5 (3.6) & {}[0.48] 1.2 (3.1) & {}[0.57] 34.6 (14.6) & \textbf{61.5}\\
\bottomrule
\end{tabular}
\end{table}

\clearpage
\section{Discussion}
The R2DT framework applies a novel parametric utility function and stopping rule in Phase I-II dose finding trials, aiming to better reflect the clinical context, including inherent uncertainty. The parameters of the utility function are informed by a structured elicitation protocol. The design has been compared to \textit{EffToxU}, an established method, which has been shown to be a special case of R2DT, and found some initial evidence that the method could lead to considerable improvement in operating characteristics. The stages of development for a new method have been likened to the 4 conventional stages of clinical trials \cite{Heinze2023}. The method is introduced from a theoretical perspective (phase I) and then applied to a small simulation study to provide proof of concept (phase II). Further work is needed to better characterise the design through more comprehensive simulation and an application of the elicitation protocol. This includes a better understanding of the number of doses, sample size, scenarios and how the parameters of R2DT reflect clinical input in elicitation.     

based upon reference dependence and attitudes to risk for both efficacy and toxicity attributes. Efficacy and toxicity utility are combined by consideration of payoffs and interaction effects to give a joint utility used to determine dosing at each stage.

The R2DT design specifies the utility function by considering the attributes of efficacy and toxicity individually. The marginal utility functions for each attribute are informed by evaluating whether an outcome is better or worse than an external standard. Attitudes toward risk differ depending on whether the outcome is perceived as a gain or a loss upon the reference. This framing is similar to that used in prospect theory \cite{Kahneman1979}, where reference dependence plays a central role. Marginal utility functions in R2DT are then combined using utility independence axioms and a simple two-parameter function to form a joint utility. Combining sigmoidal utilities in R2DT is more general, more flexible and capable of capturing true preference than existing approaches that combine simpler objective functions and limit unethical choices through admissibility rules. The intention of R2DT is to specific a utility function that better reflects both the clinical situation and unethical choices for patients simultaneously. 

The probability model specified in this paper follows that of previous work \cite{Thall2004}. All decision functions explored in this paper specify identical probability models to focus on decision theoretic elements and better understand their effects. The specification of the probability model has a large effect on the performance of different methods, particularly in terms of how likely they are to get ``stuck'' at certain doses \cite{Thall2014a}. The Bayesian decision theoretic approach, however, separates the probability and decision components \cite{Smith2010}, allowing R2DT to be applied flexibly across different settings. Modifying the probability model, such as incorporating a plateau effect \cite{Riviere2016} to better reflect modern drug behaviour, can be done without altering the utility function. Model-assisted designs have gained popularity due to their relative simplicity and favourable operating characteristics \cite{Zhang2024}. Designs such as U-BOIN \cite{Zhou2019} and BOIN12 \cite{Lin2020} use a four-outcome patient utility function to guide decisions; this function was the main comparator in this paper, although paired with the more complex EffTox logistic regression probability model. The R2DT utility function could be specified as part of a model assisted design, but further work would be needed to evaluate the merit. The motivation for using a utility function that better captures the clinical situation, as is the aim of R2DT, remains pertinent.

Any elicitation is subject to bias or error. Using appropriate methods and having an awareness of the main sources of bias ensures that a utility function is as accurate as possible \cite{Gilovich2002}.  Elicitation methods using probability lottery equivalencies are well established. To reduce bias, attribute levels  should be reasonably close together in a space that is well understood \cite{Farquhar1984}. For example considering a lottery between perfect efficacy and zero efficacy will induce bias as both levels are rarely encountered simultaneously in practice. Lottery equivalents to elicit preferences for R2DT incorporate choices made routinely in clinical practice (i.e around the reference). The impact of the reference point in the elicitation of utilities in the healthcare setting has also been investigated \cite{RodriguezMiguez2019}. Further work is ongoing to establish whether the the functional form of R2DT sufficiently captures clinical preferences through elicitation. Elicitation of the R2DT method needs a far greater understanding of utility theory and is more time consuming at the design stage than the \textit{EffToxU} design. A preliminary application of the R2DT elicitation protocol, not reported in this paper, demonstrating that while the process is more involved, it is achievable within a reasonable time frame and was understood by the clinician involved.

R2DT uses utility independence to join two separate marginal utility functions. The attribute of probabilities for both the individual utility functions could be adapted to accommodate different endpoints. Marginal continuous and time to event outcomes could be transformed into the utility scale \cite{Yuan2009}. The R2DT method would suggest that the marginal utility function for a different endpoint would be framed according to a reference point and informed by considering uncertain judgement as for the elicitation protocol in this paper. Ordinal outcomes for toxicity or efficacy would first need combining into a single measure or utility function before combining with utility independence. The R2DT utility function is evaluated in the context of sequential patient cohorts, with a decision made after each cohort. It could also be adapted for studies with alternative design elements, such as adaptive randomisation or dose-ranging trials.

\section*{Author contributions}

AH wrote the manuscript and conducted the simulation work. SRB, DW and SB contributed to the design, interpretation of the results and critically revised the manuscript. All authors read and approved the final manuscript

\section*{Acknowledgments}
This work was part of a successful PhD with funding provided by Myeloma UK

\section*{Conflict of interest}

The authors declare no potential conflict of interests.

\bibliographystyle{unsrt}
\bibliography{wileyNJD-AMA}

\appendix
\renewcommand{\thefigure}{A\arabic{figure}}
\renewcommand{\thetable}{A\arabic{table}}
\setcounter{figure}{0}
\setcounter{table}{0}

\section{Programming code\label{app:prog}}
The R code to generate all data within the manuscript is provided on a public repository: \url{https://github.com/medahala} 

\section{EffTox utility design}\label{app:EffToxU}
The EffTox utility design \cite{Thall2012, Yuan2016a} is used as a comparator to assess the R2DT in the simulation study. This section shows that the EffTox utility design can be formulated as a special case of R2DT that assumes simple risk neutral marginal utility functions. 

The EffTox Utility design specifies a discrete utility function on the four possible individual patient level outcomes, $Y=(Y_E=a,Y_T=b)$, as follows 

\begin{equation} 
u(Y_E=a,Y_T=b) = 
\begin{cases}
    K(1,1), &\text{for }a=1\text{ and }b=1\\
    K(0,0), &\text{for }a=0\text{ and }b=0\\
    K(1,0), &\text{for }a=1\text{ and }b=0\\
    K(0,1), &\text{for }a=0\text{ and }b=1\\    
\end{cases}
\end{equation}
Where $K(a, b)$ are constants to be specified. Given that utility is indifferent to linear transformations, $K(1,0)=1$ and $K(0,1)=0$ can be specified as the best and worst outcomes respectively. 
Expected utility is calculated by averaging the utility function over the chance of a state of nature (each patient outcome) happening. For the EffTox utility design the expectation is given by 
\begin{equation} 
E(u(Y(Y_E=a,Y_T=b)) = \int \sum_{a=0}^{1}\sum_{b=0}^{1} K(a, b) \pi_{a,b},
\end{equation}
where $\pi_{ab}$ represent the probability of an event happening. Assuming independence with $\pi_{11}=\pi_E\pi_T$, $\pi_{00}=(1-\pi_E)(1-\pi_T)$, $\pi_{10}=\pi_E(1-\pi_T)$, $\pi_{01}=(1-\pi_E)\pi_T$,   
and standardising with $K(0,1)=0$ and $K(1,0)=1$, the expectation equation can be rewritten as a function of $\pi_E$ and $\pi_T$:
\begin{equation}
     E(u(Y)) = E(u(\pi_E,\pi_T)) = \\
\int_\theta K(1,1)\pi_E + K(0,0)(1-\pi_T) 
+ (1 - K(0,0) - K(1,1))\pi_E(1-\pi_T) d\theta   
\end{equation}
The expected utility equation can be written as a function of the population level parameters for the probability of an event at each dose. The specific equation has been written in this form as it is analogous to the utility independence equation, Equation \ref{eq:utili_indep} with $K(1,1)=k_E$, $K(0,0)=k_T$, $u_E=\pi_E$ and $u_T=1-\pi_T$. The marginal utility functions are the identity function or the degenerate case of R2DT, $\lambda_E=\lambda_T=\alpha_{GE}=\alpha_{LE}=\alpha_{GT}=\alpha_{LT}=1$, with  $\overline{\pi}_T$ and $\overline{\pi}_E$ becoming redundant in this special case due to the normalisation function. This demonstrates that the EffTox utility design can be formulated as a special case of R2DT that assumes simple risk neutral marginal utility functions with interpretation from the perspective of population level parameters. In the paper introduction it was stated the design was indifference to decisions under uncertainty as the marginal utility functions are linear (risk neutral).

\section{Joint utility}\label{app:joint_utility}
R2DT assumes a number of conditions to define the utility function in the form $u(\pi_E,\pi_T)=f(u_E(\pi_E),u_T(\pi_T))$ with $f(\cdot)$ a linear function, $u_E$ a marginal utility function of $\pi_E$ , and $u_T$ a marginal utility function of $\pi_T$. These conditions and interpretation of additional parameters defined in $f(\cdot)$ are given in this section.  

With attributes $\pi_E$, efficacy, and $\pi_T$ toxicity,  consider a point $(e,t)$, within the domain of all possible levels $\pi_E \times \pi_T$, such that 
\begin{equation}
 0 \le e \le 1 \quad \text{and} \quad 0 \le t \le 1   
\end{equation}
Consider two conditional utility functions $u(e',\cdot)$ and $u(e'',\cdot)$ from two points $e'$ and $e''$. Defining a lottery from the conditional utility function  $u(e',\cdot)$ concerning two points $t_1$ and $t_2$ and associated certainty equivalent $\hat{t}$. We then contrast this with the certainty equivalent from the same lottery from the conditional utility function $u(e'',\cdot)$. If the certainty equivalent, $\hat{t}$ does not shift we can say that the two are strategically equivalent. 

Efficacy is utility independent of toxicity when conditional preferences for lotteries on $\pi_E$ given $\pi_T$ do not depend on the particular level of $t$. When efficacy and toxicity are mutually utility independent we can express the utility function $u(e,t)$ in a multi-linear (bilinear) form as Equation \ref{eq:utili_indep}, \cite{Keeney1971}.

The marginal utility functions $u_E$ and $u_T$ do not depend on the level of the other attribute, as per the condition of mutual utility independence, as such these are referred to as efficacy and toxicity marginal utility functions for simplicity. The constant $k_{ET}$ represents an interaction between the two attributes. A smaller sum of $k_E$ and $k_T$ would constitute a greater interaction and $k_{ET}=0$ no interaction. 

When combining two measures of consequence through a function to give a single measure of consequence as is the case here it is necessary to have an understanding of what the function is achieving; the key to this is the interaction term.

The simplest case is the independent case, this is also called additive utility independence. With additive utility independence there is no interaction term ($k_ET=0$) and the relationship between the two attributes is a simple linear payoff. There is only a single parameter that needs specifying since $k_T=1-k_E$. A small incremental increase in efficacy utility is directly proportional to a increase in toxicity utility (lower toxicity) with the magnitude of the constant dictating how much a small incremental increase in efficacy utility is worth in terms of the same increase toxicity utility. This simple payoff remains constant at all levels of efficacy and toxicity. 

\begin{figure}[ht!]
\includegraphics[width=\textwidth]{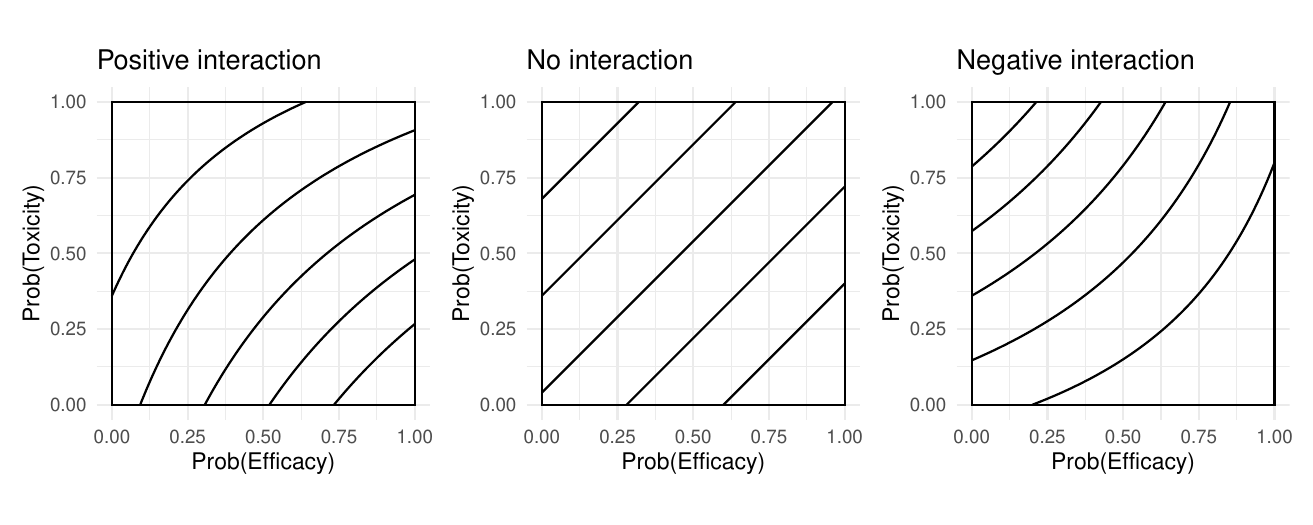}
    \caption[Example to visualise the effect of the interaction component of the joint utility function]{Example to visualise the effect of the interaction component of the joint utility function. All plots follow the utility independent relation in Equation \ref{eq:utili_indep} with simple risk neutral marginal utility functions i.e. $u_E=\pi$ and $u_T=1-\pi$. For the Positive interaction plot, $k_E=0.25$, $k_T=0.25$ and $k_{ET}=0.5$. For the no interaction plot,  $k_E=0.5$, $k_T=0.5$ and $k_{ET}=0$. For the negative interaction plot, $k_E=0.75$, $k_T=0.75$ and $k_{ET}=-0.5$. It can be seen how the slope of contour changes with utility for each of the different interactions.
    } 
    \label{fig:interaction_effect}
    \end{figure}

A positive interaction is when $k_E + k_T < 1$ and would imply that the higher the efficacy utility, the greater (more positive) the effect of toxicity utility (reduction in toxicity) on overall utility. Similarly, the higher the toxicity utility, the greater (more positive) the effect of efficacy utility on overall utility. The opposite being true of a negative value for the interaction parameter. This description of each possible interpretation for the interaction term is plotted in Figure \ref{fig:interaction_effect}. It can be seen for the plot with no interaction that the slope for toxicity with respect to efficacy is a constant at points within the joint domain. For a positive interaction the slope for toxicity with respect to efficacy is initially steep at the left hand end of the contour and reduces moving left to right. This suggests that as toxicity increases the effect of efficacy is reduced. The interpretation is synonymous with the clinical situation described for the motivating example in Section \ref{sct:intro}, with the effect of additional efficacy when there is high toxicity being minimal. A negative interaction describes the opposite to the situation in that the slope gets progressively steeper or the effect of additional efficacy becomes greater with more toxicity.   

The Figure represents a simplification of the marginal utility functions. Considering the utility with positive interaction with respect to a reference dependence and whether each attribute is a gain or loss. The interpretation is that both attributes need to be a 'gain' for the overall utility to be considered likewise. In terms of losses, if one attribute is a loss this is almost as bad as if both attributes are losses - in both cases neither would likely be suitable to treat the wider population. This is the case in oncology dose finding settings where the payoff becomes more beneficial when both attributes improve.

\section{Admissible Criteria}\label{app:stopping_rules}

The admissibility criteria used as a comparator are defined separately in relation to cut points $\overline{\pi}_{addE}$ and $\overline{\pi}_{addT}$ and evidence levels $p_E$ and $p_T$: 
\begin{align} \label{eq:admis_e}
\text{Pr}\left\{ \pi_{E}  < \overline{\pi}_{addE}~|~y \right\} & > 1-p_E \\
\label{eq:admis_t}
\text{Pr}\left\{ \pi_T >  \overline{\pi}_{addT}~|~y \right\} & > 1-p_T
\end{align}
If either criteria is met the dose will be excluded from the set $D$. If all doses meet the criteria the trial is stopped.

\section{Additional Tables and Figures}\label{app:tabs_and_figs}

\begin{table}[ht!]
\caption{\label{tab:model_params}Listing of each of the probability and fixed trial parameters for simulation study}
\centering
\begin{tabular}[t]{lll}
\toprule
Notation & Value & Interpretation\\
\midrule
${D}$ & $[20,30,40,50]$ & actual doses\\
${x}$ & $[-0.5,-0.1,0.19,0.41]$ & transformed doses\\
${x^2}$ & $[0.25,0.01,0.04,0.17] $ & square of transformed doses\\
$\alpha_T$ & $N(-3.17, 2.88)$ & toxicity intercept\\
$\beta_{1T}$ & $N(-3.56, 2.79)$ & toxicity slope\\
$\alpha_E$ & $N(0.73, 2.44)$ & efficacy intercept\\
$\beta_{1E}$ & $N(-0.11, 2.34) $ & efficacy slope\\
$\beta_{2E}$ & $N(0, 0.2) $ & efficacy squared slope\\
$\tilde{\pi}_E(D)$ & $[0.42,0.57,0.67,0.72] $ & Efficacy prior probabilities\\
$\tilde{\pi}_T(D)$ & $[0.14,0.2,0.26,0.33] $ & Toxicity prior probabilities\\
 & $[1, 1]$ & ESS toxicity and efficacy\\
 & 20 & Starting dose\\
N & 45 & Max Sample Size\\
 & 3 & Cohort Size\\
 & 2000 & Number of simulation repetitions\\
\bottomrule
\end{tabular}
\end{table}

\begin{figure}[ht!]
\includegraphics[width=0.95\textwidth]{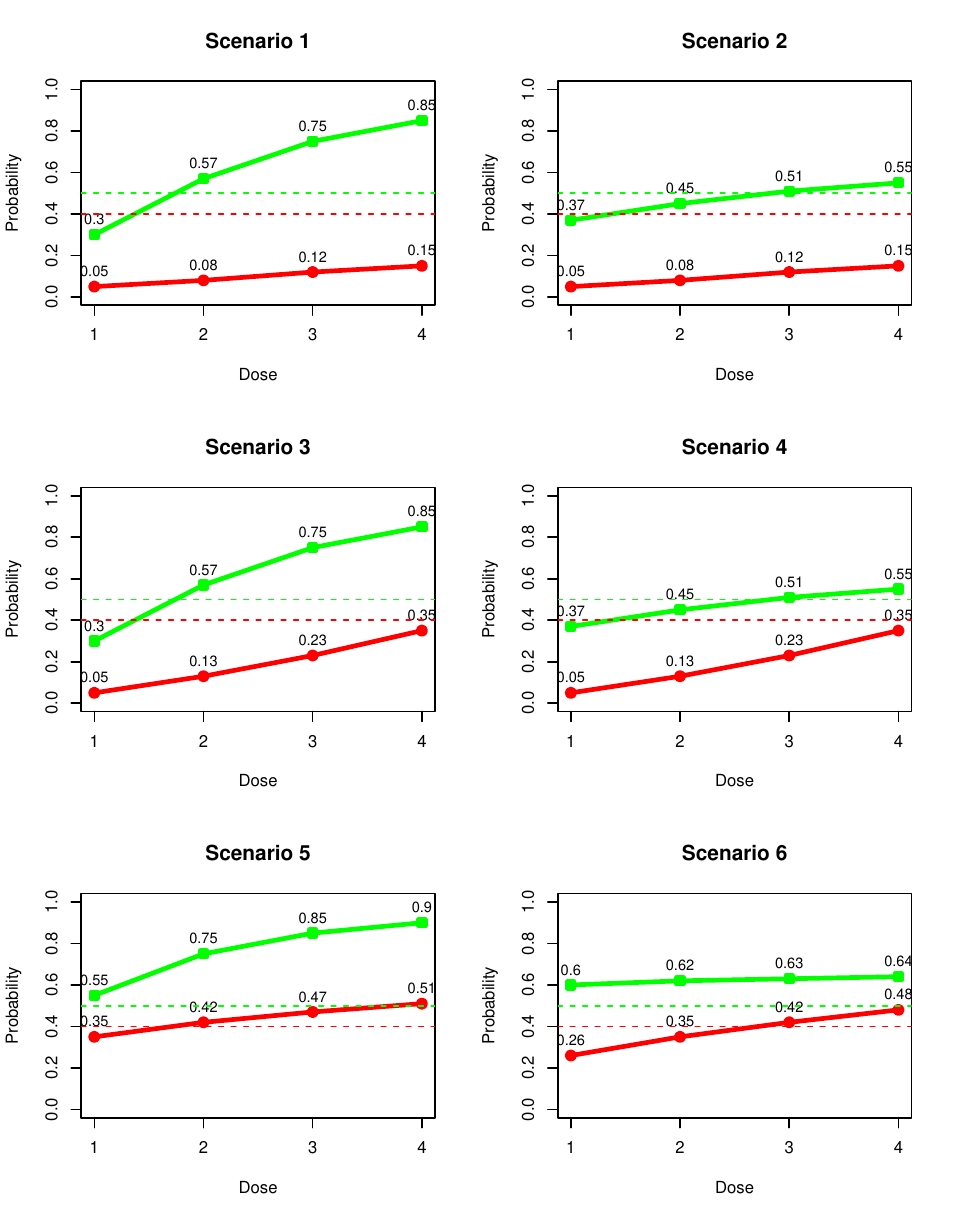}
    \caption{Plot of first six scenarios in simulation study Scenarios 1:6. Green line is the fixed probabilities for efficacy $(\tilde{\pi}_E(D))$ and red line for toxicity $(\tilde{\pi}_T(D))$. Dashed lines represent the cut points for the admissibility rules given in \textit{R2DT (1)}}
    \label{fig:scen_1_R2DT}
    \end{figure}
 \newpage   
\begin{figure}[ht!]
\includegraphics[width=\textwidth]{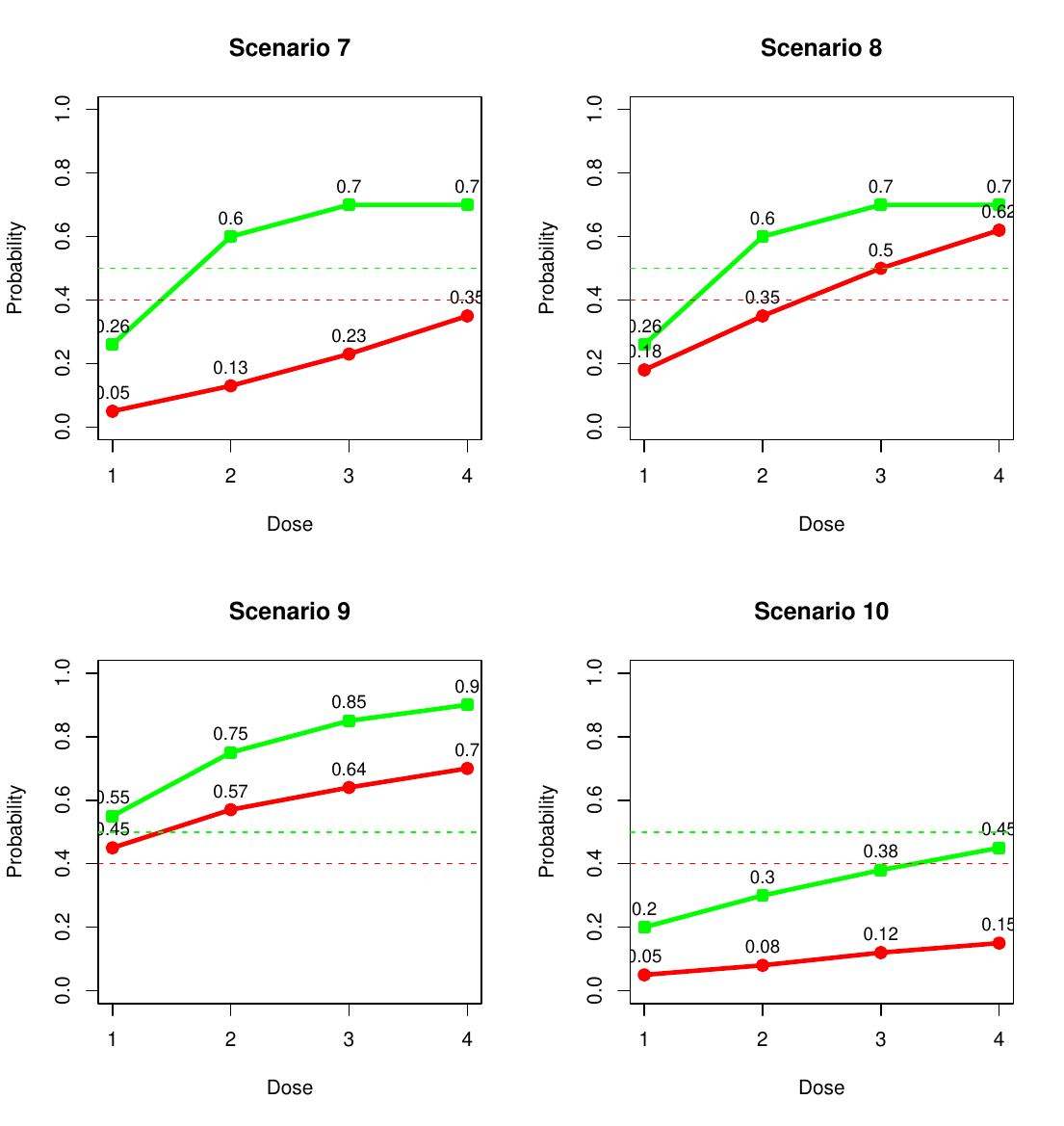}
    \caption[Plot of scenarios seven to ten in simulation study]{Scenarios 7:10. Green line is the fixed probabilities for efficacy $(\tilde{\pi}_E(D))$ and red line for toxicity $(\tilde{\pi}_T(D))$. Dashed lines represent the cut points for the admissibility rules given in \textit{R2DT (1)}}
    \label{fig:scen_2_R2DT}
    \end{figure}
\newpage
\begin{figure}[ht!]
\includegraphics[width=0.92\textwidth]{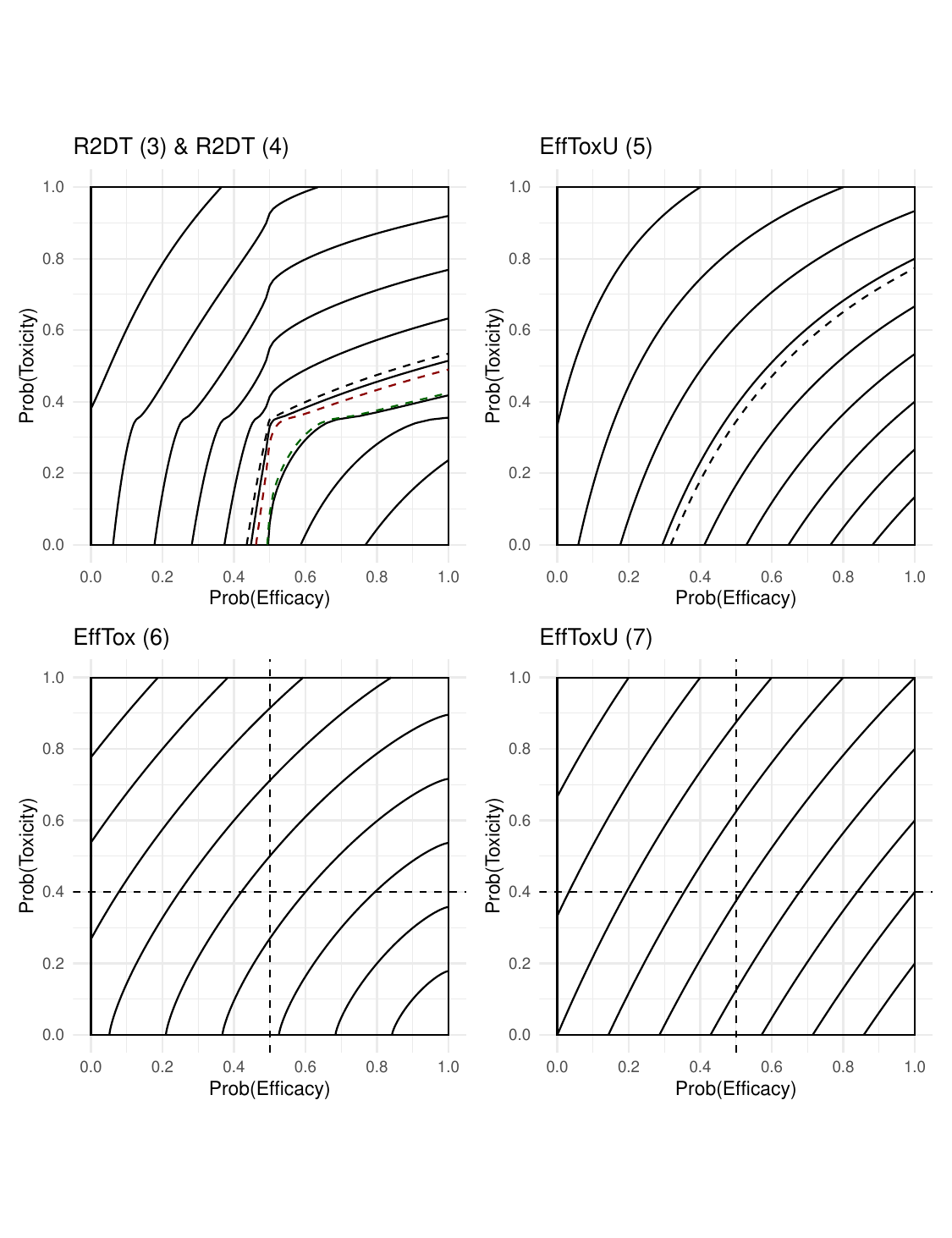}
  \caption[Contour plots for utility functions used in simulation study]{Simulation Utility Functions:
       Contours in the joint utility represent equal utility at 0.1,0.2,...,0.9 with the the point at guaranteed efficacy and no toxicity having utility of 1. Dashed lines are limits for admissibility rules. The contour plot for  \textit{R2DT (3)} and \textit{R2DT (4)} gives the stopping rule \textit{(i)} in black ($u(0.5,0.35)= 0.58$), \textit{(ii)} in red ($u(0.7,0.4)= 0.62$) and \textit{(iii} in Green ($u(0.9,0.4)= 0.69$)}
       \label{fig:scenerio_utility} 
\end{figure} 

\begin{table}[ht!]

\caption[Simulation study results applying novel stopping rule to EffToxU]{\label{tab:Table_5} Simulation study results applying novel stopping rule to \textit{EffToxU}: data of form: [utility at scenario probability $(\pi_E, \pi_T)$] percentage selection (average number of patients treated). Percentage of trials with no dose selected abbreviated to NDS. Bold indicates optimal dose or whether trial
should recommend not selecting a dose (and stop early)}
\centering
\begin{tabular}[t]{llllll}
\toprule
\multicolumn{1}{c}{ } & \multicolumn{4}{c}{Dose (mg/kg)} & \multicolumn{1}{c}{ } \\
\cmidrule(l{3pt}r{3pt}){2-5}
\multicolumn{1}{l}{Method} & \multicolumn{1}{c}{20} & \multicolumn{1}{c}{30} & \multicolumn{1}{c}{40} & \multicolumn{1}{c}{50} & \multicolumn{1}{c}{NDS}\\
\midrule
\addlinespace[0.3em]
\multicolumn{6}{c}{\textbf{Scenario 1} $(\pi_E, \pi_T)$}\\
\multicolumn{1}{c}{\hspace{1em}} & \multicolumn{1}{c}{(0.3, 0.05)} & \multicolumn{1}{c}{(0.57, 0.08)} & \multicolumn{1}{c}{(0.75, 0.12)} & \multicolumn{1}{c}{(0.85, 0.15)} & \multicolumn{1}{c}{}\\
\midrule
\hspace{1em}EffToxU (2) & {}[0.39] 1.5 (4.8) & {}[0.60] 4.1 (5.3) & {}[0.72] 3.8 (4.1) & {}[\textbf{0.77}] 90.4 (30.8) & 0.2\\
\hspace{1em}EffToxU (5) & {}[0.39] 3.2 (5.3) & {}[0.60] 4 (5.2) & {}[0.72] 3.6 (4.1) & {}[\textbf{0.77}] 89.1 (30.4) & 0\\
\addlinespace[0.3em]
\multicolumn{6}{c}{\textbf{Scenario 2} $(\pi_E, \pi_T)$}\\
\multicolumn{1}{c}{\hspace{1em}} & \multicolumn{1}{c}{(0.37, 0.05)} & \multicolumn{1}{c}{(0.45, 0.08)} & \multicolumn{1}{c}{(0.51, 0.12)} & \multicolumn{1}{c}{(0.55, 0.15)} & \multicolumn{1}{c}{}\\
\midrule
\hspace{1em}EffToxU (2) & {}[0.45] 15.3 (11.6) & {}[0.50] 5.8 (5.9) & {}[0.53] 6.5 (4.2) & {}[\textbf{0.55}] 65 (21.8) & 7.4\\
\hspace{1em}EffToxU (5) & {}[0.45] 24.9 (13.2) & {}[0.50] 6.2 (6) & {}[0.53] 5 (3.9) & {}[\textbf{0.55}] 63.5 (21.6) & 0.5\\
\addlinespace[0.3em]
\multicolumn{6}{c}{\textbf{Scenario 3} $(\pi_E, \pi_T)$}\\
\multicolumn{1}{c}{\hspace{1em}} & \multicolumn{1}{c}{(0.3, 0.05)} & \multicolumn{1}{c}{(0.57, 0.13)} & \multicolumn{1}{c}{(0.75, 0.23)} & \multicolumn{1}{c}{(0.85, 0.35)} & \multicolumn{1}{c}{}\\
\midrule
\hspace{1em}EffToxU (2) & {}[0.39] 1.2 (4.8) & {}[0.57] 9 (6.7) & {}[\textbf{0.65}] 29.2 (10.2) & {}[0.64] 60.1 (23.1) & 0.5\\
\hspace{1em}EffToxU (5) & {}[0.39] 2.9 (5.2) & {}[0.57] 8.8 (6.6) & {}[\textbf{0.65}] 28.3 (10.1) & {}[0.64] 60 (23) & 0.1\\
\addlinespace[0.3em]
\multicolumn{6}{c}{\textbf{Scenario 4} $(\pi_E, \pi_T)$}\\
\multicolumn{1}{c}{\hspace{1em}} & \multicolumn{1}{c}{(0.37, 0.05)} & \multicolumn{1}{c}{(0.45, 0.13)} & \multicolumn{1}{c}{(0.51, 0.23)} & \multicolumn{1}{c}{(0.55, 0.35)} & \multicolumn{1}{c}{}\\
\midrule
\hspace{1em}EffToxU (2) & {}[0.45] 16.2 (11.8) & {}[0.48] 13.3 (7.5) & {}[\textbf{0.48}] 21.3 (7.4) & {}[0.45] 40.5 (16.4) & 8.7\\
\hspace{1em}EffToxU (5) & {}[0.45] 28.2 (14.1) & {}[0.48] 13.5 (7.5) & {}[\textbf{0.48}] 17.6 (6.7) & {}[0.45] 38.4 (16) & 2.2\\
\addlinespace[0.3em]
\multicolumn{6}{c}{\textbf{Scenario 5} $(\pi_E, \pi_T)$}\\
\multicolumn{1}{c}{\hspace{1em}} & \multicolumn{1}{c}{(0.55, 0.35)} & \multicolumn{1}{c}{(0.75, 0.42)} & \multicolumn{1}{c}{(0.85, 0.47)} & \multicolumn{1}{c}{(0.9, 0.51)} & \multicolumn{1}{c}{}\\
\midrule
\hspace{1em}EffToxU (2) & {}[0.45] 14.8 (7.6) & {}[0.54] 26.9 (11.4) & {}[0.56] 15.6 (7.3) & {}[\textbf{0.56}] 34.2 (17.2) & 8.5\\
\hspace{1em}EffToxU (5) & {}[0.45] 8.1 (6.4) & {}[0.54] 20.6 (10.4) & {}[0.56] 16.6 (7.8) & {}[\textbf{0.56}] 54.6 (20.4) & 0\\
\addlinespace[0.3em]
\multicolumn{6}{c}{\textbf{Scenario 6} $(\pi_E, \pi_T)$}\\
\multicolumn{1}{c}{\hspace{1em}} & \multicolumn{1}{c}{(0.6, 0.26)} & \multicolumn{1}{c}{(0.62, 0.35)} & \multicolumn{1}{c}{(0.63, 0.42)} & \multicolumn{1}{c}{(0.64, 0.48)} & \multicolumn{1}{c}{}\\
\midrule
\hspace{1em}EffToxU (2) & {}[\textbf{0.53}] 39.1 (14.8) & {}[0.49] 24.6 (12.8) & {}[0.46] 9.8 (5.8) & {}[0.44] 24.3 (11.2) & 2.1\\
\hspace{1em}EffToxU (5) & {}[\textbf{0.53}] 38 (14.6) & {}[0.49] 22.8 (12.6) & {}[0.46] 8.6 (5.6) & {}[0.44] 29.8 (12) & 0.8\\
\addlinespace[0.3em]
\multicolumn{6}{c}{\textbf{Scenario 7} $(\pi_E, \pi_T)$}\\
\multicolumn{1}{c}{\hspace{1em}} & \multicolumn{1}{c}{(0.26, 0.05)} & \multicolumn{1}{c}{(0.6, 0.13)} & \multicolumn{1}{c}{(0.7, 0.23)} & \multicolumn{1}{c}{(0.7, 0.35)} & \multicolumn{1}{c}{}\\
\midrule
\hspace{1em}EffToxU (2) & {}[0.36] 0.9 (4.6) & {}[0.59] 11.9 (7.4) & {}[\textbf{0.61}] 27.9 (9.4) & {}[0.54] 58.6 (23.4) & 0.8\\
\hspace{1em}EffToxU (5) & {}[0.36] 2.2 (4.9) & {}[0.59] 11.8 (7.4) & {}[\textbf{0.61}] 27.1 (9.3) & {}[0.54] 58.6 (23.3) & 0.4\\
\addlinespace[0.3em]
\multicolumn{6}{c}{\textbf{Scenario 8} $(\pi_E, \pi_T)$}\\
\multicolumn{1}{c}{\hspace{1em}} & \multicolumn{1}{c}{(0.26, 0.18)} & \multicolumn{1}{c}{(0.6, 0.35)} & \multicolumn{1}{c}{(0.7, 0.5)} & \multicolumn{1}{c}{(0.7, 0.62)} & \multicolumn{1}{c}{}\\
\midrule
\hspace{1em}EffToxU (2) & {}[0.32] 3.9 (6.3) & {}[\textbf{0.48}] 50.8 (14.4) & {}[0.46] 26.5 (10.6) & {}[0.39] 11.8 (12.3) & 7\\
\hspace{1em}EffToxU (5) & {}[0.32] 5.9 (6.5) & {}[\textbf{0.48}] 28.3 (10.8) & {}[0.46] 24.1 (10.2) & {}[0.39] 35.7 (16.5) & 5.9\\
\addlinespace[0.3em]
\multicolumn{6}{c}{\textbf{Scenario 9} $(\pi_E, \pi_T)$}\\
\multicolumn{1}{c}{\hspace{1em}} & \multicolumn{1}{c}{(0.55, 0.45)} & \multicolumn{1}{c}{(0.75, 0.57)} & \multicolumn{1}{c}{(0.85, 0.64)} & \multicolumn{1}{c}{(0.9, 0.7)} & \multicolumn{1}{c}{}\\
\midrule
\hspace{1em}EffToxU (2) & {}[0.40] 29.4 (12.3) & {}[0.45] 13.2 (8.9) & {}[0.45] 2 (4.3) & {}[0.43] 2.9 (8.6) & \textbf{52.5}\\
\hspace{1em}EffToxU (5) & {}[0.40] 13.7 (8.6) & {}[0.45] 20.8 (9.3) & {}[\textbf{0.45}] 13.4 (6.6) & {}[0.43] 48.9 (19.9) & 3.2\\
\addlinespace[0.3em]
\multicolumn{6}{c}{\textbf{Scenario 10} $(\pi_E, \pi_T)$}\\
\multicolumn{1}{c}{\hspace{1em}} & \multicolumn{1}{c}{(0.2, 0.05)} & \multicolumn{1}{c}{(0.3, 0.08)} & \multicolumn{1}{c}{(0.38, 0.12)} & \multicolumn{1}{c}{(0.45, 0.15)} & \multicolumn{1}{c}{}\\
\midrule
\hspace{1em}EffToxU (2) & {}[0.31] 1.5 (6.1) & {}[0.38] 0.9 (3.7) & {}[0.43] 1.5 (3.6) & {}[0.47] 51.1 (21.7) & \textbf{44.9}\\
\hspace{1em}EffToxU (5) & {}[0.31] 6.1 (8) & {}[0.38] 1.9 (3.9) & {}[0.43] 2.4 (3.5) & {}[\textbf{0.47}] 77.8 (26.4) & 11.8\\
\bottomrule
\end{tabular}
\end{table}

\newpage

\newpage
\begin{table}[ht!]

\caption[Simulation study results sensitivity of EffToxU specification]{\label{tab:Table_3}Sensitivity of EffToxU: data of form: [utility at scenario probability $(\pi_E, \pi_T)$] percentage selection (average number of patients treated). Percentage of trials with no dose selected abbreviated to NDS. Bold indicates optimal dose or whether trial
should recommend not selecting a dose (and stop early)}
\centering
\scalebox{0.9}{
\begin{tabular}[t]{llllll}
\toprule
\multicolumn{1}{c}{ } & \multicolumn{4}{c}{Dose (mg/kg)} & \multicolumn{1}{c}{ } \\
\cmidrule(l{3pt}r{3pt}){2-5}
\multicolumn{1}{l}{Method} & \multicolumn{1}{c}{20} & \multicolumn{1}{c}{30} & \multicolumn{1}{c}{40} & \multicolumn{1}{c}{50} & \multicolumn{1}{c}{NDS}\\
\midrule
\addlinespace[0.3em]
\multicolumn{6}{c}{\textbf{Scenario 1} $(\pi_E, \pi_T)$}\\
\multicolumn{1}{c}{\hspace{1em}} & \multicolumn{1}{c}{(0.3, 0.05)} & \multicolumn{1}{c}{(0.57, 0.08)} & \multicolumn{1}{c}{(0.75, 0.12)} & \multicolumn{1}{c}{(0.85, 0.15)} & \multicolumn{1}{c}{}\\
\midrule
\hspace{1em}EffToxU (2) & {}[0.39] 1.5 (4.8) & {}[0.60] 4.1 (5.3) & {}[0.72] 3.8 (4.1) & {}[\textbf{0.77}] 90.4 (30.8) & 0.2\\
\hspace{1em}EffTox (6) & {}[0.55] 1.4 (4.8) & {}[0.71] 4.2 (5.4) & {}[0.81] 3.5 (4.1) & {}[\textbf{0.85}] 90.6 (30.6) & 0.2\\
\hspace{1em}EffToxU (7) & {}[0.49] 1.5 (4.8) & {}[0.67] 4.4 (5.3) & {}[0.77] 2.2 (3.7) & {}[\textbf{0.82}] 91.6 (31.2) & 0.2\\
\addlinespace[0.3em]
\multicolumn{6}{c}{\textbf{Scenario 2} $(\pi_E, \pi_T)$}\\
\multicolumn{1}{c}{\hspace{1em}} & \multicolumn{1}{c}{(0.37, 0.05)} & \multicolumn{1}{c}{(0.45, 0.08)} & \multicolumn{1}{c}{(0.51, 0.12)} & \multicolumn{1}{c}{(0.55, 0.15)} & \multicolumn{1}{c}{}\\
\midrule
\hspace{1em}EffToxU (2) & {}[0.45] 15.3 (11.6) & {}[0.50] 5.8 (5.9) & {}[0.53] 6.5 (4.2) & {}[\textbf{0.55}] 65 (21.8) & 7.4\\
\hspace{1em}EffTox (6) & {}[0.59] 15.4 (11.7) & {}[0.64] 5.9 (6.1) & {}[0.66] 5.9 (4) & {}[\textbf{0.68}] 65.6 (21.8) & 7.2\\
\hspace{1em}EffToxU (7) & {}[0.54] 15.3 (11.4) & {}[0.58] 6 (6) & {}[0.61] 6.6 (4.2) & {}[\textbf{0.62}] 65 (21.9) & 7.1\\
\addlinespace[0.3em]
\multicolumn{6}{c}{\textbf{Scenario 3} $(\pi_E, \pi_T)$}\\
\multicolumn{1}{c}{\hspace{1em}} & \multicolumn{1}{c}{(0.3, 0.05)} & \multicolumn{1}{c}{(0.57, 0.13)} & \multicolumn{1}{c}{(0.75, 0.23)} & \multicolumn{1}{c}{(0.85, 0.35)} & \multicolumn{1}{c}{}\\
\midrule
\hspace{1em}EffToxU (2) & {}[0.39] 1.2 (4.8) & {}[0.57] 9 (6.7) & {}[\textbf{0.65}] 29.2 (10.2) & {}[0.64] 60.1 (23.1) & 0.5\\
\hspace{1em}EffTox (6) & {}[0.55] 1 (4.8) & {}[0.69] 8.4 (6.6) & {}[\textbf{0.76}] 31 (10.6) & {}[0.75] 59.2 (22.9) & 0.4\\
\hspace{1em}EffToxU (7) & {}[0.49] 1.1 (4.8) & {}[0.64] 7.3 (6.2) & {}[0.72] 19.9 (8) & {}[\textbf{0.73}] 71.2 (25.9) & 0.5\\
\addlinespace[0.3em]
\multicolumn{6}{c}{\textbf{Scenario 4} $(\pi_E, \pi_T)$}\\
\multicolumn{1}{c}{\hspace{1em}} & \multicolumn{1}{c}{(0.37, 0.05)} & \multicolumn{1}{c}{(0.45, 0.13)} & \multicolumn{1}{c}{(0.51, 0.23)} & \multicolumn{1}{c}{(0.55, 0.35)} & \multicolumn{1}{c}{}\\
\midrule
\hspace{1em}EffToxU (2) & {}[0.45] 16.2 (11.8) & {}[0.48] 13.3 (7.5) & {}[\textbf{0.48}] 21.3 (7.4) & {}[0.45] 40.5 (16.4) & 8.7\\
\hspace{1em}EffTox (6) & {}[0.59] 16.2 (11.8) & {}[\textbf{0.62}] 14.1 (7.7) & {}[0.62] 21.3 (7.5) & {}[0.60] 39.7 (16.1) & 8.6\\
\hspace{1em}EffToxU (7) & {}[0.54] 17.2 (11.9) & {}[0.56] 13.2 (7.4) & {}[\textbf{0.56}] 18.4 (6.8) & {}[0.54] 43 (17) & 8.2\\
\addlinespace[0.3em]
\multicolumn{6}{c}{\textbf{Scenario 5} $(\pi_E, \pi_T)$}\\
\multicolumn{1}{c}{\hspace{1em}} & \multicolumn{1}{c}{(0.55, 0.35)} & \multicolumn{1}{c}{(0.75, 0.42)} & \multicolumn{1}{c}{(0.85, 0.47)} & \multicolumn{1}{c}{(0.9, 0.51)} & \multicolumn{1}{c}{}\\
\midrule
\hspace{1em}EffToxU (2) & {}[0.45] 14.8 (7.6) & {}[0.54] 26.9 (11.4) & {}[0.56] 15.6 (7.3) & {}[\textbf{0.56}] 34.2 (17.2) & 8.5\\
\hspace{1em}EffTox (6) & {}[0.60] 14.1 (7.6) & {}[0.67] 27.7 (11.3) & {}[\textbf{0.69}] 15.3 (7.5) & {}[0.69] 34.4 (17.1) & 8.5\\
\hspace{1em}EffToxU (7) & {}[0.54] 12.4 (7.2) & {}[0.64] 26.2 (10.7) & {}[0.67] 14.5 (7.1) & {}[\textbf{0.69}] 38.6 (18.6) & 8.3\\
\addlinespace[0.3em]
\multicolumn{6}{c}{\textbf{Scenario 6} $(\pi_E, \pi_T)$}\\
\multicolumn{1}{c}{\hspace{1em}} & \multicolumn{1}{c}{(0.6, 0.26)} & \multicolumn{1}{c}{(0.62, 0.35)} & \multicolumn{1}{c}{(0.63, 0.42)} & \multicolumn{1}{c}{(0.64, 0.48)} & \multicolumn{1}{c}{}\\
\midrule
\hspace{1em}EffToxU (2) & {}[\textbf{0.53}] 39.1 (14.8) & {}[0.49] 24.6 (12.8) & {}[0.46] 9.8 (5.8) & {}[0.44] 24.3 (11.2) & 2.1\\
\hspace{1em}EffTox (6) & {}[\textbf{0.66}] 39.1 (14.8) & {}[0.63] 24.8 (12.8) & {}[0.61] 9.2 (5.8) & {}[0.58] 25 (11.3) & 2\\
\hspace{1em}EffToxU (7) & {}[\textbf{0.61}] 40.6 (15.7) & {}[0.59] 21.4 (11.8) & {}[0.56] 9.7 (5.5) & {}[0.54] 26 (11.6) & 2.1\\
\addlinespace[0.3em]
\multicolumn{6}{c}{\textbf{Scenario 7} $(\pi_E, \pi_T)$}\\
\multicolumn{1}{c}{\hspace{1em}} & \multicolumn{1}{c}{(0.26, 0.05)} & \multicolumn{1}{c}{(0.6, 0.13)} & \multicolumn{1}{c}{(0.7, 0.23)} & \multicolumn{1}{c}{(0.7, 0.35)} & \multicolumn{1}{c}{}\\
\midrule
\hspace{1em}EffToxU (2) & {}[0.36] 0.9 (4.6) & {}[0.59] 11.9 (7.4) & {}[\textbf{0.61}] 27.9 (9.4) & {}[0.54] 58.6 (23.4) & 0.8\\
\hspace{1em}EffTox (6) & {}[0.52] 0.9 (4.6) & {}[0.71] 12.7 (7.5) & {}[\textbf{0.73}] 27.4 (9.4) & {}[0.68] 58.4 (23.4) & 0.7\\
\hspace{1em}EffToxU (7) & {}[0.46] 0.9 (4.7) & {}[0.66] 10.4 (6.8) & {}[\textbf{0.69}] 20.2 (7.8) & {}[0.64] 67.7 (25.5) & 0.8\\
\addlinespace[0.3em]
\multicolumn{6}{c}{\textbf{Scenario 8} $(\pi_E, \pi_T)$}\\
\multicolumn{1}{c}{\hspace{1em}} & \multicolumn{1}{c}{(0.26, 0.18)} & \multicolumn{1}{c}{(0.6, 0.35)} & \multicolumn{1}{c}{(0.7, 0.5)} & \multicolumn{1}{c}{(0.7, 0.62)} & \multicolumn{1}{c}{}\\
\midrule
\hspace{1em}EffToxU (2) & {}[0.32] 3.9 (6.3) & {}[\textbf{0.48}] 50.8 (14.4) & {}[0.46] 26.5 (10.6) & {}[0.39] 11.8 (12.3) & 7\\
\hspace{1em}EffTox (6) & {}[0.49] 4.3 (6.6) & {}[\textbf{0.62}] 50.3 (14.4) & {}[0.60] 26 (10.4) & {}[0.54] 12 (12.4) & 7.2\\
\hspace{1em}EffToxU (7) & {}[0.42] 4.7 (6.4) & {}[\textbf{0.57}] 42.8 (12.9) & {}[0.57] 30.8 (10.3) & {}[0.52] 14.6 (14.2) & 7.2\\
\addlinespace[0.3em]
\multicolumn{6}{c}{\textbf{Scenario 9} $(\pi_E, \pi_T)$}\\
\multicolumn{1}{c}{\hspace{1em}} & \multicolumn{1}{c}{(0.55, 0.45)} & \multicolumn{1}{c}{(0.75, 0.57)} & \multicolumn{1}{c}{(0.85, 0.64)} & \multicolumn{1}{c}{(0.9, 0.7)} & \multicolumn{1}{c}{}\\
\midrule
\hspace{1em}EffToxU (2) & {}[0.40] 29.4 (12.3) & {}[0.45] 13.2 (8.9) & {}[0.45] 2 (4.3) & {}[0.43] 2.9 (8.6) & \textbf{52.5}\\
\hspace{1em}EffTox (6) & {}[0.55] 30 (12.5) & {}[0.59] 13.2 (8.8) & {}[0.60] 1.6 (4.3) & {}[0.58] 3 (8.6) & \textbf{52.1}\\
\hspace{1em}EffToxU (7) & {}[0.50] 28.9 (11.9) & {}[0.57] 15 (9.1) & {}[0.59] 2 (4.5) & {}[0.59] 3.1 (8.8) & \textbf{51}\\
\addlinespace[0.3em]
\multicolumn{6}{c}{\textbf{Scenario 10} $(\pi_E, \pi_T)$}\\
\multicolumn{1}{c}{\hspace{1em}} & \multicolumn{1}{c}{(0.2, 0.05)} & \multicolumn{1}{c}{(0.3, 0.08)} & \multicolumn{1}{c}{(0.38, 0.12)} & \multicolumn{1}{c}{(0.45, 0.15)} & \multicolumn{1}{c}{}\\
\midrule
\hspace{1em}EffToxU (2) & {}[0.31] 1.5 (6.1) & {}[0.38] 0.9 (3.7) & {}[0.43] 1.5 (3.6) & {}[0.47] 51.1 (21.7) & \textbf{44.9}\\
\hspace{1em}EffTox (6) & {}[0.49] 1.7 (6.1) & {}[0.54] 0.7 (3.7) & {}[0.58] 1.8 (3.6) & {}[0.61] 51.1 (21.7) & \textbf{44.8}\\
\hspace{1em}EffToxU (7) & {}[0.42] 1.4 (6.1) & {}[0.48] 0.8 (3.7) & {}[0.52] 1.8 (3.6) & {}[0.56] 51.2 (21.7) & \textbf{44.9}\\
\bottomrule
\end{tabular}
}
\end{table}

\end{document}